\documentclass[preprint, review, number]{elsarticle}
 \usepackage[utf8]{inputenc}
 \usepackage{amsmath, amsfonts, amsthm}
 \usepackage{duckuments}
 \usepackage{graphicx}
 \usepackage[T1]{fontenc}
 \usepackage{svg}
 \usepackage{enumitem,amssymb}
 \usepackage{ulem}
 \newlist{todolist}{itemize}{2}
 \setlist[todolist]{label=$\square$}

\usepackage[resetlabels]{multibib}
\newcites{SI}{References}

\usepackage{todonotes}
 \usepackage{appendix}
 \usepackage{cases}
 \usepackage{empheq}
 
\usepackage{hyperref}

\usepackage[margin=2cm]{geometry}

\begin{document}

\begin{frontmatter}

\title{Phenotypic switching mechanisms determine the structure of cell migration into extracellular matrix under the `go-or-grow' hypothesis}

\author[1]{Rebecca~M.~Crossley} %\corref{cor1}}
\ead{rebecca.crossley@maths.ox.ac.uk}
\author[2]{Kevin J. Painter}
\ead{kevin.painter@polito.it}
\author[3]{Tommaso Lorenzi}
\ead{tommaso.lorenzi@polito.it}
\author[1]{Philip~K.~Maini}
\ead{philip.maini@maths.ox.ac.uk}
\author[1]{Ruth E. Baker}
\ead{ruth.baker@maths.ox.ac.uk}

\cortext[cor1]{Corresponding author.}

\affiliation[1]{organization={Mathematical Institute},
addressline={University of Oxford},
postcode={OX2 6GG},
city={Oxford},
country={United Kingdom}}

\affiliation[2]{organization={Dipartimento di Scienze, Progetto e Politiche del Territorio (DIST)},
addressline={Politecnico di Torino},
postcode={10129},
city={Torino},
country={Italy}}

\affiliation[3]{organization={Department of Mathematical Sciences ``G. L. Lagrange''},
addressline={Politecnico di Torino},
postcode={10129},
city={Torino},
country={Italy}}

%%%%%%%% Abstract
\begin{abstract}
A fundamental feature of collective cell migration is phenotypic heterogeneity which{, for example,} influences tumour progression and relapse. 
While current mathematical models often consider discrete phenotypic structuring of the cell population, in-line with the `go-or-grow' hypothesis \cite{hatzikirou2012go, stepien2018traveling}, they regularly overlook the role that the {environment may play} in determining the cells' {phenotype} during migration.
Comparing a previously studied {volume-filling model for a} homogeneous population {of generalist cells that can proliferate, move and degrade} extracellular matrix (ECM) \cite{crossley2023travelling} to a novel {model for a} heterogeneous population {comprising two} distinct sub-populations {of specialist cells that can either move and degrade ECM or proliferate}, this study explores how different {hypothetical} phenotypic switching mechanisms affect the speed and structure {of the {invading} cell populations}. 
%Both models take into account the volume-filling affects of cells and ECM.
Through a continuum model derived from its individual-based counterpart, insights into the influence of the ECM and the impact of {phenotypic} switching on migrating cell populations emerge. 
Notably, specialist cell {populations that cannot switch phenotype} show reduced invasiveness compared to generalist cell {populations}, while implementing different {forms of} switching significantly alters the structure of migrating cell fronts.
{This key result suggests that the structure of an invading cell population could be used to infer the underlying mechanisms governing phenotypic switching.}
\end{abstract}

%%%%%%%% Keywords

\end{frontmatter}

%%%%%%%%%%%% Intro
\section{Introduction}
Phenotypic heterogeneity profoundly impacts tumour behaviour and is a hallmark feature driving post-treatment recurrence \cite{summerbell2020epigenetically}. 
Collective cell migration, which {can be} crucial in understanding {various} {stages of} tumour progression, often involves distinct cell phenotypes with varying motility and proliferative capacities \cite{giese1996dichotomy, giese1996migration}. 
Mathematical approaches to studying these processes often {use} models formulated as reaction-diffusion equations with phenotypic structuring \cite{macfarlane2022impact, lorenzi2022invasion}, where cells are assumed to undergo random, undirected movement and grow logistically to some maximum capacity, similar to the classical Fisher-KPP model \cite{fisher_wave_1937, kolmogorov1937study}. 

It is often observed experimentally that individual cells are either proliferative or motile \cite{konen2017image}, but not both \cite{campbell2021cooperation}, representing a trade-off known as the `go-or-grow' hypothesis  \cite{kolbe2020modeling, pham2012density}. 
Numerous mathematical models have been {proposed} to study the migration-proliferation dichotomy of a phenotypically structured population of cells, such as those that simulate glioblastoma growth \cite{gerlee2012impact}. 
{Models of this nature} consider various phenotypic switching mechanisms, derive analytical expressions for the minimum travelling wave speed of the migrating front \cite{stepien2018traveling, gerlee2016travelling, curtin2020speed, tursynkozha2023traveling}, or include more complex non-linear diffusion terms \cite{conte2021mathematical}. 
Crucially, however, cell phenotypes and their functions are fundamentally dependent on external cues, such as contact with neighbouring cells or interactions with the extracellular matrix (ECM) {--} the highly complex network of proteins and other macromolecules that cells reside within {--} which reduces the available space for migration and provides structure and chemical cues to guide migration \cite{werb1997ecm, cox2006new, bloom2014influence}. 
Despite the increasing evidence for phenotypically structured cell populations, few studies have included the role of the ECM in models for cell migration under the go-or-grow hypothesis, to consider how it might affect the phenotypic structure of invading fronts and their speed. 
Furthermore, despite several experimental results supporting the existence of leader and follower cell {sub-populations} during collective cell migration \cite{zoeller2019genetic, mclennan2015neural}, {{other} findings {do not support such hypotheses} \cite{vittadello2020examining, GARAY20133094}, possibly {due to the use of} different cell types or experimental conditions.}

To investigate the role of the ECM further, this work compares a model of a {homogeneous population of generalist cells that can proliferate, move and degrade ECM} to a model of a heterogeneous {population comprising two sub-populations of specialist cells that can either move and degrade ECM or proliferate.}
This continuum model for two distinct cell phenotypes and ECM {dynamics} is derived from first principles (as the limit of an underlying individual-based model) to accurately account for individual cell mechanisms at the population level.
A number of different possible phenotypic switching mechanisms are considered, including random switching and both cell and ECM dependent switches in various forms, to explore their impact on the speed and phenotypic structure of {invading cell populations}.

\subsection{Layout of the paper}
{In Section~\ref{model}, we begin by} describing the underlying assumptions of the models we study. 
We start by reviewing the details of the model for a population of homogeneous generalist cells in one spatial dimension that {was} first derived in \cite{crossley2023travelling}. 
Subsequently, we extend this model to a heterogeneous population comprising two distinct sub-populations of specialist cells that can either move and degrade ECM or proliferate, in order to introduce phenotypic heterogeneity into the cell population (see Section~\ref{2model} and~\ref{model_deriv_app}).
In Section~\ref{results}, the solutions of these models are studied numerically for a variety of different phenotypic switching functions. 
We find that specialist cell population{s} with the ability to switch phenotype {may invade faster} than a generalist cell population, and that the choice of phenotypic switching mechanism drastically impacts the phenotypic structure of migrating cell fronts, such that the leading cell sub-population differs between switching functions. 
{The phenotypic structure of the invading cell population could potentially therefore be used to predict the underlying switching mechanism.}
In Section~\ref{discuss}{,} we discuss these findings, the possible applications and potential avenues for future research. 

%%%%%%%%%%%%%%% Math Model
\section{Mathematical {m}odels}\label{model}
It is {well}-known that cells move in response to gradients in local cell volume fractions, and in response to nearby environmental features, such as the ECM, by haptotaxis{,} for example \cite{alberts2017molecular}. 
Many mathematical models have been developed that include non-linear terms to describe these processes, and non-local reaction terms to describe {the} proliferation of cells {to} fill the surrounding available space \cite{gerisch2008mathematical, browning2019bayesian}. 

To invade into surrounding healthy tissues, {many tumours} must overcome physical barriers to migration{,} such as the ECM. 
In order to do this, tumour cells have developed mechanisms such as the ability to remodel, reorient and degrade elements of the ECM \cite{lee2017local, winkler2020concepts} through the production of specific matrix degrading enzymes, such as matrix metalloproteases (MMPs), that act in very close proximity to their cell of origin before decaying \cite{stetler1993tumor, chaplain2005mathematical, kessenbrock2010matrix}. 
Since the timescale of ECM degradation is much longer than the timescale of intermediate processes, such as MMP {decay}, we employ {the} simplifying assumption that cells directly degrade the ECM \cite{perumpanani1999extracellular}.
In this work, we {focus on investigating the simplest possible problem -- that of {the role of} phenotypic heterogeneity in cell invasion into an ECM that is devoid of cells.}

\begin{figure}[htbp]
    \centering
    \includegraphics[width=\linewidth]{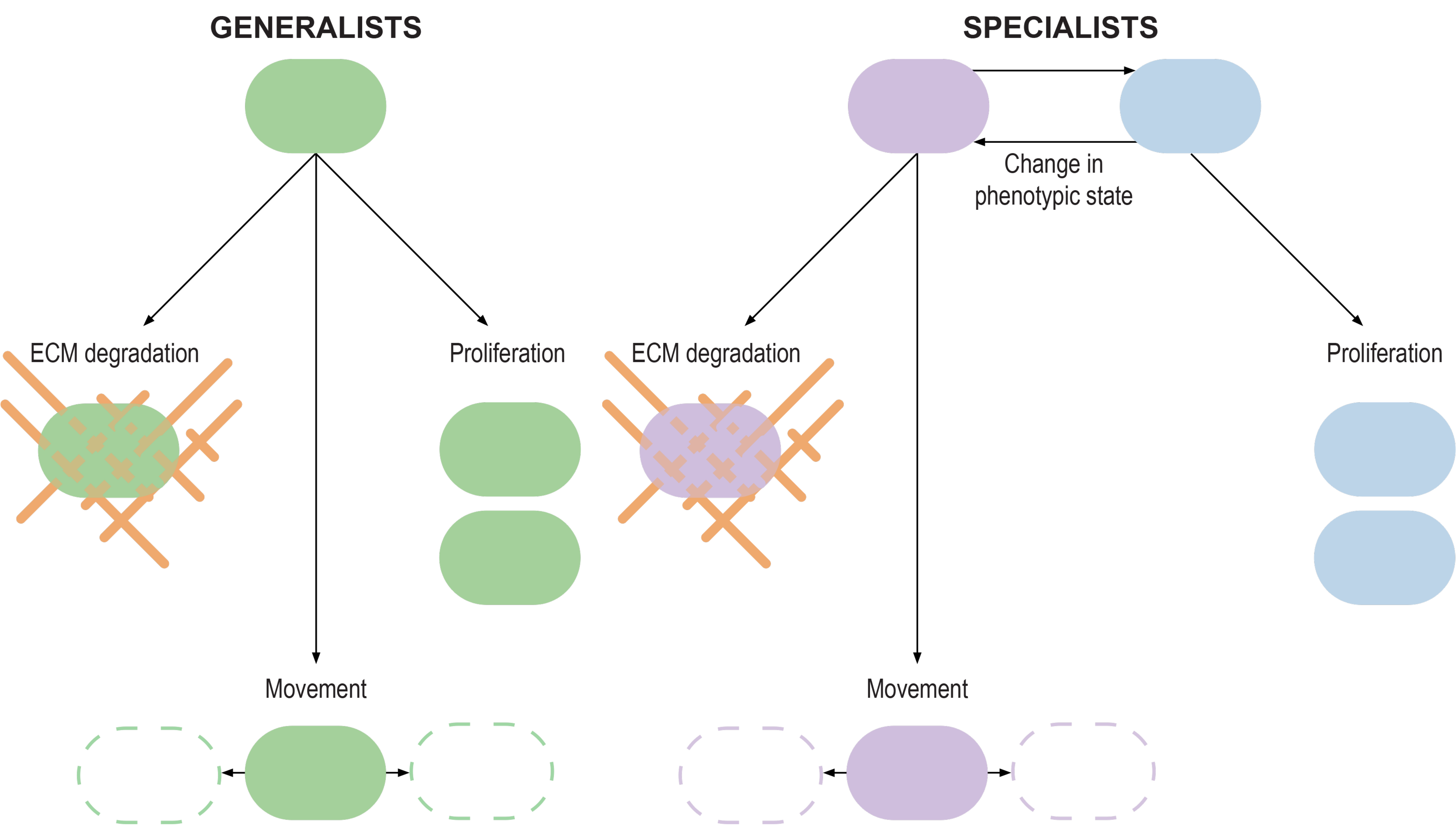}
    \caption{{Schematic} representation of the model for a homogeneous generalist population of cells, and
    the model for a heterogeneous specialist population of cells.}
    \label{fig:cartoon_schem_model}
\end{figure}

This study focuses on cell movement and proliferation, as restricted by volume filling assumptions that {entail} both of these processes {being} limited by the presence of other surrounding cells and ECM, and on the degradation of ECM by direct contact with the cells.   
We begin by presenting two deterministic, continuum models for cell migration into {the} ECM that have been derived by coarse-graining underlying individual-based models to {give rise to} {a} corresponding population-level description (see schematic in Fig.~\ref{fig:cartoon_schem_model}). 
The first model {considers} a homogeneous generalist cell population invading into the ECM, as is often studied in {standard} models for collective cell migration \cite{painter2009modelling}, and builds on similar models for cell migration into the ECM studied in \cite{el2021travelling, colson2021travelling}. 
To further extend this previous work, we introduce a second model describing a population of specialist cells consisting of two distinct phenotypes, in line with evidence supporting the existence of separate proliferating and migrating populations \cite{konen2017image}. 
This model extends those presented in \cite{pham2012density, chauviere2010model, saut2014multilayer} to include the ECM and its degradation by cells, as well as volume-filling effects. 
We then compare the resulting structure of {travelling wave} solutions {of the models} and investigate differences between the migrating cell population distributions and {invasion} speeds.

\subsection{{A model for a homogeneous generalist cell population invading into the ECM}}\label{1model}
We first consider a homogeneous generalist population of cells that (under volume-filling assumptions) is motile, proliferative and degrades the ECM. 
Previous models considered cell migration into {the} ECM without volume-filling assumptions \cite{el2021travelling, colson2021travelling}{; these models comprised} two coupled differential equations with non-linear cross-dependent diffusion and logistic growth. 
{The differential equation model for generalist cells considered here} describes the evolution of cell and ECM densities under volume-filling assumptions, where the movement and proliferation of cells is reduced in higher volume fraction regions{.
In \cite{crossley2023travelling}, this model} was derived from an underlying one-dimensional, on-lattice, individual-based model, and its travelling wave solutions {were} studied. 

\begin{figure}[h!]
    \centering
    \includegraphics[width=\linewidth]{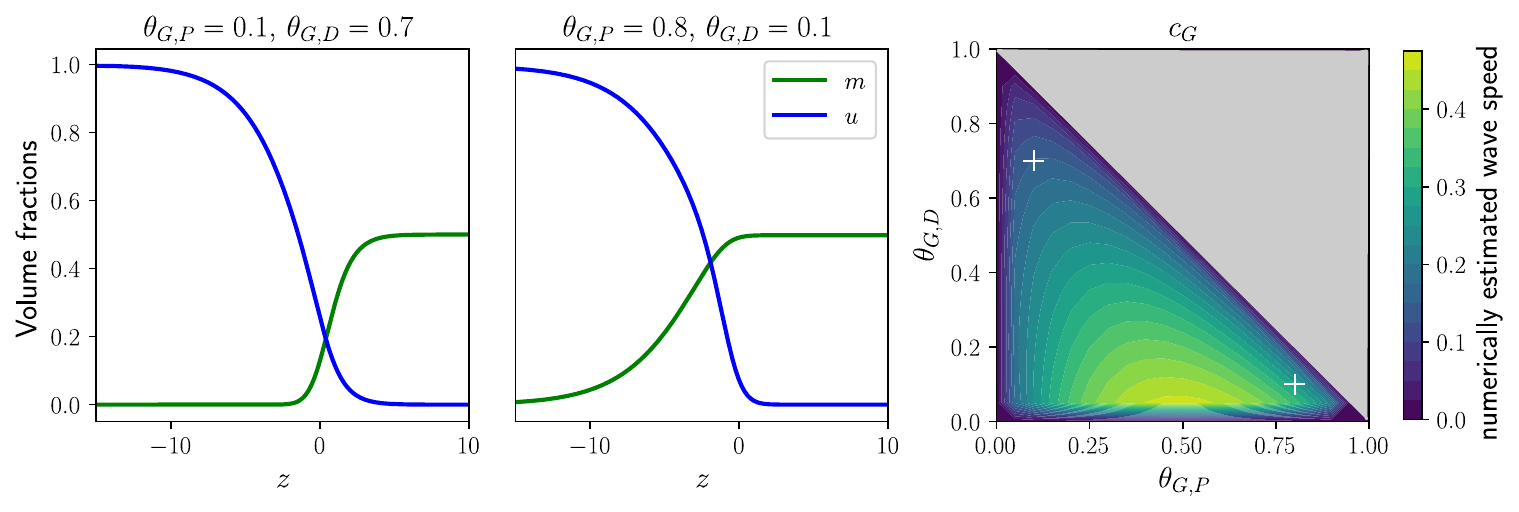}
    \caption{Plots of the travelling wave {profile} of the system~\eqref{gen_weight_u}-\eqref{gen_weight_m} subject to the initial conditions for the cells as in Eq.~\eqref{IC_u} and {for} the ECM as in Eq.~\eqref{IC_m} for different values of $\theta_{G,D}$ and $\theta_{G,P}$ {and translated into the travelling wave co-ordinate, $z=x-c_Gt$,} where $c_G$ is the numerically estimated travelling wave speed. The far right shows the contour plot of the numerically observed travelling wave speeds as we vary $\theta_{G,D}$ and $\theta_{G,P}.$ The initial ECM volume fraction ahead of the cells is $m_0=0.5$, the ECM degradation rate is $\lambda=1$, and the width of the {region initially occupied by} migrating cells is $\alpha=1$ across all simulations. For more information regarding the numerical methods used see \ref{app_methods}.}
    \label{fig:gen}
\end{figure}

To motivate later comparisons with a heterogeneous cell population, we re-introduce this model here with non-dimensional weightings of the cells towards proliferation, $\theta_{G,P}\in[0,1]$, degradation of ECM, $\theta_{G,D}\in[0,1],$ and movement, $(1-\theta_{G,D}-\theta_{G,P})\in[0,1],$ {that distribute a cells'} {weighting} across {different} functions.  
The non-dimensional volume fractions of the {generalist} cell population and {corresponding} ECM are denoted as $u_{G}(x,t)$ and $m_{G}(x,t)$, respectively, and their dynamics are governed by the following system{:}
\begin{align}
    \frac{\partial {u_G}}{\partial t}&= (1-\theta_{G,P}-\theta_{G,D})\frac{\partial}{\partial {x}}\Bigg[ \big(1-u_G-{m_G}\big)\frac{\partial u_G}{\partial {x}}+u_G \frac{\partial }{\partial {x}}\big({u_G+{m_G}}\big)\Bigg]+ \theta_{G,P}u_G(1-u_G-m_G), \label{gen_weight_u} \\
    \frac{\partial {m_{G}}}{\partial {t}}  &=-\theta_{G,D} {\lambda_G} {m_{G}}{u_G}, \label{gen_weight_m} 
\end{align}
%2col version...
% \begin{align}
%     \frac{\partial {u_G}}{\partial t}&= (1-\theta_{G,P}-\theta_{G,D})\frac{\partial}{\partial {x}}\Bigg[ \big(1-u_G-{m}\big)\frac{\partial u_G}{\partial {x}}\nonumber \\ &\qquad\qquad\qquad\qquad\qquad\qquad+u_G \frac{\partial }{\partial {x}}\big({u_G+{m}}\big)\Bigg]\nonumber \\ &\qquad\qquad+ \theta_{G,P}u_G(1-u_G-m), \label{gen_weight_u} \\
%     \frac{\partial {m_{G}}}{\partial {t}}  &=-\theta_{G,D} {\lambda_G} {m_{G}}{u_G}, \label{gen_weight_m} 
% \end{align}
where $x\in\mathbb{R}$ and $t\geq0$.
The first term inside the square brackets on the right-hand side of Eq.~\eqref{gen_weight_u} models the {undirected} movement {(i.e., diffusion)} of the generalist cells, where this movement is prevented by the presence of other cells and ECM. 
The second term inside the square brackets models movement of the cells down the {gradient of the} total volume fraction of both cells and ECM, $u_G+m_G$, whereas the {last} (reaction) term describes proliferation of cells, which is assumed to be logistic up to a carrying capacity in the total of the cell and ECM volume fractions {(non-dimensionalised to {unity})}. 
The parameter $\lambda_G\in\mathbb{R}_{+}$ is the rescaled ECM degradation rate and when $\lambda_G=0$ this model may simplify to a Fisher-KPP model with appropriately rescaled parameters \cite{crossley2023travelling}.
We employ the following initial conditions:
\begin{equation}
    u_{G}(x,0)=\begin{cases}1, \qquad &$if$ \qquad x<\alpha , \\ 0, \qquad &$if$ \qquad x\geq\alpha, \end{cases}\label{IC_u} 
\end{equation}
\begin{align}
    m_{G}(x,0)= \begin{cases} 
    0, \qquad &$if$ \qquad x<\alpha, \\
    m_0, \qquad &$if$ \qquad x\geq\alpha, \label{IC_m}
    \end{cases} 
\end{align}
%\begin{align}
%    0&\leq u_G(x,0), \, m_G(x,0)\leq1, \\
%    0&\leq u_G(x,0)+m_G(x,0)\leq1,
%\end{align}
where  $m_0\in[0,1)$ corresponds to the volume fraction of ECM ahead of the cells, and $\alpha\in\mathbb{R}_{+}$ defines the width of the {region initially occupied by the} cells.
We complement this model with {the following} boundary conditions:
$u_G$ and $\partial u_G / \partial x \to 0$ as $x\to\infty.$
Previous studies show that{, under these conditions,} travelling wave solutions can be observed, whose speed{s} depend on the initial volume fraction of ECM ahead of the invading wave of cells and the ECM degradation rate \cite{crossley2023travelling, colson_travelling-wave_2021}. 
Examples of these solutions, and their numerically estimated travelling wave speeds as the parameters $\theta_{G,D}$ and $\theta_{G,P}$ vary, are shown in Figure~\ref{fig:gen}.

\subsection{{A model for a heterogeneous specialist cell population invading into the ECM}}\label{2model}
Having introduced {the} model for a {homogeneous generalist population of cells} migrating into the ECM, we now extend it to investigate the impact of phenotypic heterogeneity. 

To do this, we consider two different cell types, whose {properties} follow the well-studied go-or-grow {hypothesis} \cite{ hatzikirou2012go, giese2003cost}.
We introduce a {discrete} variable $p\in\{1,2\}$ that represents the cell phenotypic state. 
Cells in the phenotypic state $p=1$ are able to degrade the ECM and are motile but cannot proliferate, whereas cells in the phenotypic state $p=2$ are able to proliferate but do not degrade ECM or move.
The volume fraction of cells in phenotypic state $p\in\{1,2\}$ at time ${t}\geq0$ is denoted by ${u_p}({x},{t})$ and the volume fraction of ECM at $t\geq0$ is denoted {by} ${m}({x},{t})$.
{The population model is obtained through coarse-graining an individual-based model (see \ref{model_deriv_app}) and, following a non-dimensionalisation (see \ref{NDapp}){,} is given by} the following system:
\begin{align}
\frac{\partial {u_1}}{\partial {t}} &=(1-\theta_{S,D})\frac{\partial}{\partial {x}}\Bigg[ \big(1-u_1-u_2-m\big)\frac{\partial {u}_1}{\partial {x}}+{u_1} \frac{\partial }{\partial {x}}\big(u_1+u_2+m\big)\Bigg] +{{u}_2}{\gamma_{21}(u_1, u_2, m)}-{u}_1{\gamma_{12}(u_1, u_2, m)},\label{intro_u1_ND} \\
\frac{\partial {u_2}}{\partial {t}}&= \theta_{S,P}{{u}_2}\big(1-u_1-u_2-m\big)-{{u}_2}{\gamma_{21}(u_1, u_2, m)}+{u}_1{\gamma_{12}(u_1, u_2, m)},  \label{intro_u2_ND} \\
\frac{\partial {m}}{\partial {t}}  &=-\theta_{S,D}\lambda {m}{u_1}, \label{intro_m_ND}
\end{align}
% for 2col..
% \begin{align}
% \frac{\partial {u_1}}{\partial {t}} &=(1-\theta_{S,D})\frac{\partial}{\partial {x}}\Bigg[ \big(1-u_1-u_2-m\big)\frac{\partial {u}_1}{\partial {x}}\nonumber \\ &\qquad\qquad\qquad\qquad\qquad\qquad+{u_1} \frac{\partial }{\partial {x}}\big(u_1+u_2+m\big)\Bigg] \nonumber \\ &\qquad\qquad+{{u}_2}{\gamma_{21}(u_1, u_2, m)}-{u}_1{\gamma_{12}(u_1, u_2, m)},\label{intro_u1_ND} \\
% \frac{\partial {u_2}}{\partial {t}}&= {{u}_2}\big(1-u_1-u_2-m\big)\nonumber \\ &\qquad\qquad-{{u}_2}{\gamma_{21}(u_1, u_2, m)}+{u}_1{\gamma_{12}(u_1, u_2, m)}, \\
% \frac{\partial {m}}{\partial {t}}  &=-\theta_{S,D}\lambda {m}{u_1}, \label{intro_m_ND}
% \end{align}
where $x\in\mathbb{R}$ and $t\in\mathbb{R}_{+}.$ 
Here{,} ${\lambda}\in\mathbb{R}_{+}$ is the rescaled rate of ECM degradation by cells in phenotypic state 1, whilst $\theta_{S,D}\in[0,1]$ describes the weighting of cells in phenotypic state 1 towards degrading ECM. 
{Similarly,} $\theta_{S,P}$ describes the weighting of cells in phenotypic state 2 towards proliferation{, where we set $\theta_{S,P}=1$ for the duration of this study,} and 
\begin{equation*}
    \gamma_{12}:\mathbb{R}_{+}^3\rightarrow\mathbb{R}_{+} \qquad \text{and} \qquad \gamma_{21}:\mathbb{R}_{+}^3\rightarrow\mathbb{R}_{+},
\end{equation*}
are the non-dimensional phenotypic switching functions{, where $\gamma_{12}$ is the phenotypic switching function from state 1 to 2 (and vice versa {for $\gamma_{21}$}).}

Similar to the generalist model, the first term inside the square brackets in {Eq.~\eqref{intro_u1_ND} captures undirected movement (i.e., diffusion) of the cells in phenotypic state 1, {inhibited} by the presence of other cells and ECM, and the second term inside the square brackets describes the movement of cells in phenotypic state 1 down the gradient in the total volume fraction of cells and ECM. 
The first term {on the right-hand side of} Eq.~\eqref{intro_u2_ND} describes the growth of cells in phenotypic state 2{,} as limited by the presence of other cells and ECM{.
The single} term on the right-hand side of Eq.~\eqref{intro_m_ND} describes degradation of ECM by cells in phenotypic state 1. 
We consider a number of phenotypic switching functions {that incorporate different types of behaviour with a view to} understand{ing} their impact on the speed of cell {invasion as well as {on} the structure of the invading wave}. 

We assume boundary conditions {of the form} $u_1\to0$ and $\partial u_1/\partial x\to0$ as $x\to\infty$. 
For consistency with the volume-filling assumptions of the model, we also initially assume that
\begin{equation}
    u_	{p}(x,0)=\begin{cases}0.5, \qquad &$if$ \qquad x<\alpha , \\ 0, \qquad &$if$ \qquad x\geq\alpha, \end{cases}\label{IC_up} 
\end{equation}
{for $p\in\{1,2\}.$}
The initial conditions for the ECM volume fraction, $m(x,0)$, are prescribed by Eq.~\eqref{IC_m}, where $\alpha$ and $m_0$ have the same interpretation as in Section~\ref{1model}.  
{Furthermore, we note here that the system~\eqref{intro_u1_ND}-\eqref{intro_m_ND} and the system~\eqref{gen_weight_u}-\eqref{gen_weight_m} are solved numerically on the domain $x\in[0,L]$, where $L$ is chosen differently between simulations to be sufficiently large such that convergence in the travelling wave speed is observed, and the boundary conditions do not have an effect (see~\ref{app_methods}).
For more information on the method for the numerical calculation of the travelling wave speed, $c$, for each simulation, see~\ref{app_methods}, and note that the solutions in Figures~\ref{fig:GvS1},~\ref{fig:dist_sketch}~and~\ref{fig:varys} are translated into the travelling wave co-ordinate $z=x-ct$.}

\subsubsection{Phenotypic switching functions}\label{PS_sec}
The phenotypic switching functions we consider, $\gamma_{12}(u_1, u_2, m)$ and $\gamma_{21}(u_1, u_2, m)$, are listed in Table~\ref{tab:PStab}.
The first is constant switching, at rate $s\in\mathbb{R}_{+}$, between the two sub-populations. 
{It is important to note that} when $s=0$ there is no switching between the phenotypic states{; in this case} {the model does not permit} travelling wave solutions and invasion is not observed. 
The second, ECM-dependent phenotypic switching{, entails cells} switching from phenotypic state 1 (2) to phenotypic state 2 (1) {at a rate that} decreases (increases) linearly with ECM volume fraction, {and} describe{s} a higher rate of switching to {the ECM} degrading {phenotypic state} in regions of higher ECM volume fractions.
{The third, s}pace-dependent phenotypic switching{,} is defined such that the rate of switching from phenotypic state 1 (2) to phenotypic state 2 (1) increases (decreases) with {the} available space, $1-u_1-u_2-m$.
{Finally,} cell-dependent phenotypic switching {assumes that} only the total} cell volume fraction impact{s} switching, {with the ECM playing no {direct} role in {driving} phenotypic switching.} 
{Note that, since $0\leq u_1+u_2+m\leq1$, all the switching functions given in Table~\ref{tab:PStab} are non-negative.}
\begin{table*}[htbp]
\begin{center}
\begin{tabular}{|c|c|c|}
    \hline
    Name & $\gamma_{12}(u_1, u_2, m)$ & $\gamma_{21}(u_1, u_2, m)$\\
    \hline\hline
    Constant & $s$ & $s$ \\
    ECM-dependent  & $s (1-m)$ & $s m$ \\
    Space-dependent  & $s(1-u_1-u_2-m)$ & $s(u_1+u_2+m)$ \\
    Cell-dependent  & $s(1-u_1-u_2)$ & $s(u_1+u_2)$ \\
    \hline
\end{tabular}
\end{center}
\caption{Table listing the phenotypic switching functions {we consider}.}
\label{tab:PStab}
\end{table*}

%%%%%%%%%%%%% Results
\section{Results}\label{results}%sti;l
\subsection{Is the speed of migration impacted by the introduction of phenotypic switching?}

In reality, many different cell types are known to {co-operate} to create robust migration, often {through} the {emergence} of leader and follower cell phenotypes.
For example, in neural crest cell migration, the leader and follower phenotypes generate streams of {invading} cells \cite{mclennan2015neural, mclennan2012multiscale, martinson2023dynamic}, whereas during angiogenesis, tip and stalk cells aid in the branching process \cite{carmeliet2011molecular}. 
{To investigate whether robust cell invasion can be observed in a phenotypically heterogeneous specialist population, we explore the dynamics of Eqs.~\eqref{intro_u1_ND}-\eqref{intro_m_ND}, first in the case of constant switching, at rate $s>0.$}

\begin{figure*}[h!]
    \centering
    \includegraphics[width=\linewidth]{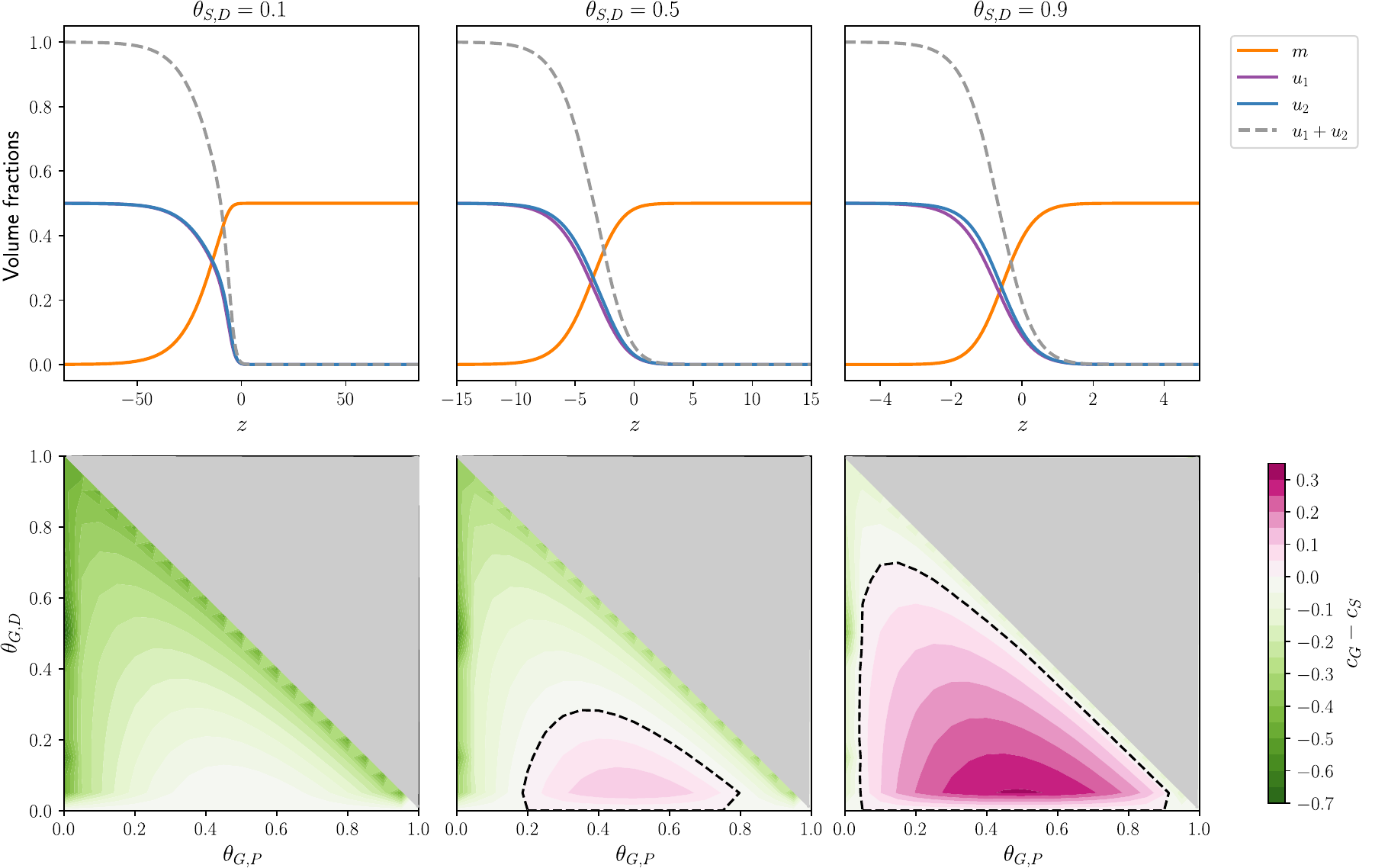}
    \caption{
    The top row shows {plots of} the travelling wave {profile} of the system \eqref{intro_u1_ND}-\eqref{intro_m_ND} subject to the initial conditions for the cells as in Eq.~\eqref{IC_up} and {for} the ECM as in Eq.~\eqref{IC_m} for different values of $\theta_{S,D}$ {for the case of constant switching (see Table~\ref{tab:PStab}) and translated into the travelling wave co-ordinate, $z=x-c_St$, where $c_S$ is the numerically estimated travelling wave speed.} The bottom row shows contour plots displaying the difference between the numerically observed travelling wave speed of the homogeneous generalist population, system~\eqref{gen_weight_u}-\eqref{gen_weight_m} subject to the initial conditions~\eqref{IC_u}-\eqref{IC_m}, and the {heterogeneous specialist population,} system \eqref{intro_u1_ND}-\eqref{intro_m_ND} subject to the initial conditions for the cells as in Eq.~\eqref{IC_up} and for the ECM as in Eq.~\eqref{IC_m}. The regions coloured in pink display the parameter regimes where the difference between the travelling wave speeds, $c_{{G}}-c_{{S}}$, is positive, meaning generalist cells {invade} faster than specialists, and regions are {coloured in} green when this difference is negative (i.e., $c_{S}> c_{G}$). The dashed black line is plotted at $c_{G}=c_{S}.$ The initial ECM volume fraction ahead of the cells is $m_0=0.5$, the ECM degradation rate is $\lambda=1$, the switching rate is $s=1$, and the width of the region initially occupied by migrating cells is $\alpha=1$ across all simulations. For more information regarding the numerical methods used see \ref{app_methods}.}
    \label{fig:GvS1}
\end{figure*}

In this case, travelling wave solutions can be observed in the ECM and cell volume fractions{, where the two sub-populations are well mixed along the invading front} (see the top row of {Fig.~\ref{fig:GvS1}}). 
The bottom row of Fig.~\ref{fig:GvS1} shows the difference between the numerically estimated travelling wave speed of a homogeneous generalist population, $c_{G}${,} and a heterogeneous specialist population, $c_{S}$, such that green indicates regions of parameter space where the travelling wave speed of the heterogeneous cell population exceeds that of the homogeneous counterpart.  
{The numerically estimated travelling wave speeds $c_{G}$ and $c_{S}$ are computed as described in \ref{app_methods}.}
The maximum observed speed {from either cell population} in these model simulations was 0.5, when $\theta_{G,P}=0.5,\, \theta_{G,D}=0$ {and $m_0<1$}, and thus the differences between the travelling wave speeds of the {homogeneous and heterogeneous} populations are observed to be of the same order of magnitude as the numerically estimated travelling wave speeds.
We find that specialist cell populations with constant, phenotypic switching have a faster travelling wave speed in all cases except when generalists {heavily} weight their abilities towards cell motility, rather than ECM degradation.
{However, it is important to note that the} maximum possible travelling wave speed {for} {the two models} is {the same} (see Supplementary Material~\ref{supp:max}). 

%%%%env dep
\subsection{Does environmentally-dependent phenotypic switching change the speed or structure of migrating fronts?}
In reality, cells are able to sense their environment and neighbouring cells, {which} can both provide cues for directed migration.
Variations in the surrounding cells and environment can also cause phenotypic changes within cells that affect their behaviour \cite{deregibus2007endothelial} and thus we extend {our study of} the heterogeneous specialist population of cells to consider the impact of ECM-, space- and cell-dependent phenotypic switching functions{,} as defined in Table~\ref{tab:PStab}. 

\begin{figure*}[htbp]
    \centering
    \includegraphics[scale=0.6]{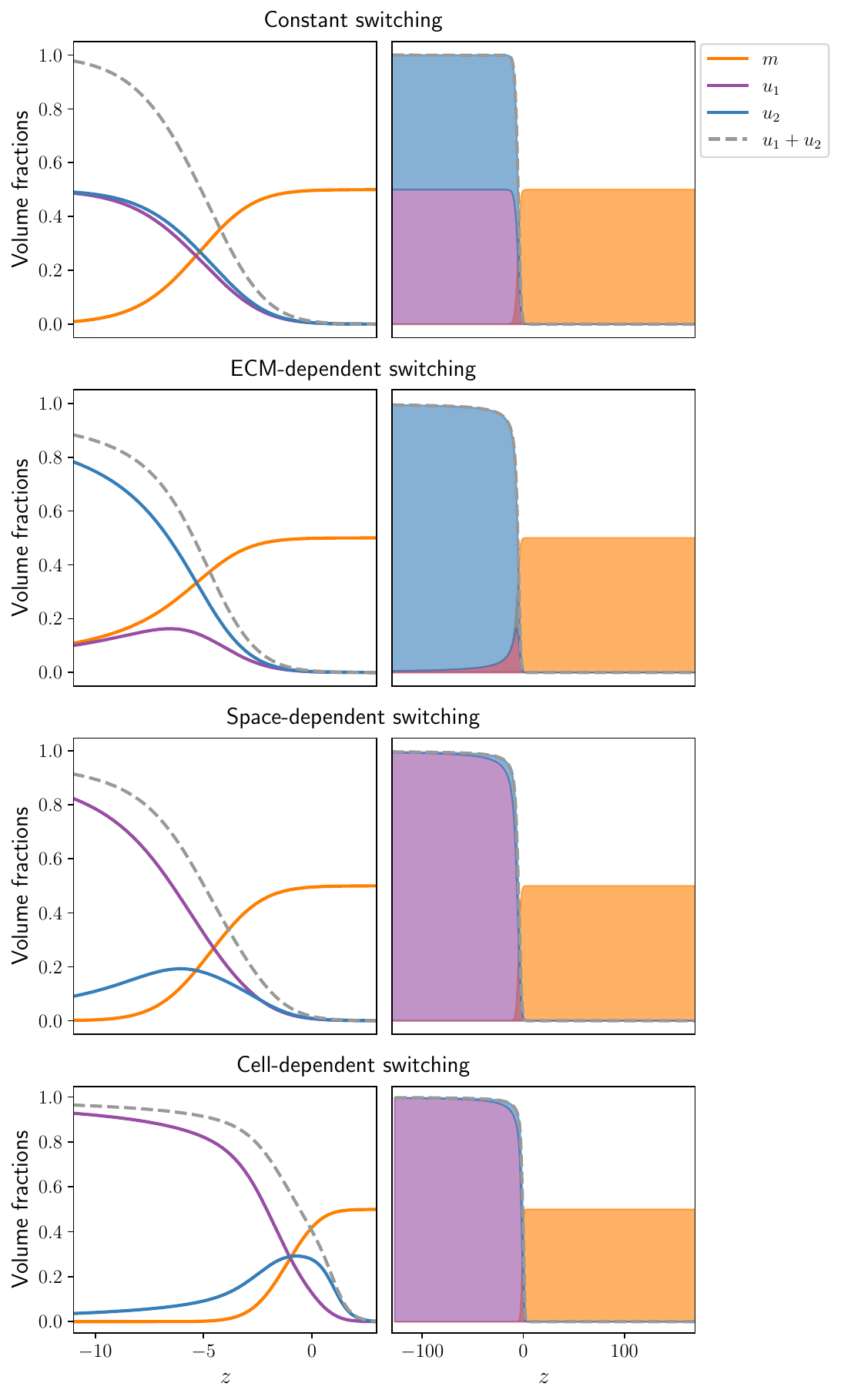}
    \caption{Plots showing the proportion of cells in each phenotypic state and their position in the travelling wave when simulating the system~\eqref{intro_u1_ND}-\eqref{intro_m_ND} subject to the initial conditions for the cells as in Eq.~\eqref{IC_up} and {for} the ECM as in Eq.~\eqref{IC_m} {and translated into the travelling wave co-ordinate, $z=x-ct$,} {where $c$ is the numerically estimated travelling wave speed,} for each of the phenotypic switching functions in Table~\ref{tab:PStab}. In the first column, we show zoomed in profiles at the front of the travelling wave. In the second column, we show the full travelling wave profile, with each constituent shaded. In all plots, the initial ECM volume fraction ahead of the cells is $m_0=0.5$ and the ECM degradation rate is $\lambda=1$. The width of {the region initially invaded by} migrating cells is $\alpha=1$, the weighting of specialists towards degradation is $\theta_{S,D}=0.5$ and the switching rate for all functions is $s=1$. {The insets in the column on the left are zoomed in on the travelling wave front.} For more information regarding the numerical methods used see \ref{app_methods}. }
    \label{fig:dist}
\end{figure*}

\paragraph{ECM-dependent switching}
We first consider the scenario where cells are able to sense the volume fraction of surrounding ECM {which then} influences the rate of phenotypic switching. 
In this case, we find that spatial heterogeneity appears within the travelling wave front (see Fig.~\ref{fig:dist}). 
Instead of a {well-}mixed population of cells that degrade and proliferate throughout the {invading} wave (as is observed for constant phenotypic switching between heterogeneous specialist populations, or for homogeneous generalist populations) we find that the cells in phenotypic state 1 (i.e.{, the} ECM degraders) concentrate at the {wave} front in the form of a travelling pulse, whereas the bulk {of the invading wave} is filled by a travelling front of proliferative cells in phenotypic state 2.  

\paragraph{Space-dependent switching}
In line with the volume-filling principles underlying this model {(i.e.{,} the fact} that cells are unable to move or proliferate in a region that has no available space{)}, we introduce phenotypic switching from phenotypic state 1 ({ECM} degrader) to phenotypic state 2 (proliferator) at an increasing rate as available space increases (see Table~\ref{tab:PStab}).  
By inspecting Fig.~\ref{fig:dist} it is clear that the bulk of the travelling wave consists of cells in phenotypic state 1, whereas cells in phenotypic state 2 concentrate at the migrating front, {which} is opposite {to what is observed} for ECM-dependent switching.
As a result, there is increased ECM degradation {due to a} larger proportion of cells in phenotypic state 1, {and we see} a sharper transition between $m=0$ and $m=m_0$ in the travelling wave as compared to constant{,} or ECM-dependent{,} switching (see Fig.~\ref{fig:dist}).

\paragraph{Cell-dependent switching}
When cells change phenotypic state {according to} the cell-dependent phenotypic switching function defined in Table~\ref{tab:PStab}, a qualitatively similar cell distribution is observed {as} in the space-dependent switching case, with proliferating cells (phenotypic state 2) at the migrating front and ECM-degrading cells (phenotypic state 1) in the bulk.
Subtle differences {between the travelling wave profile for ECM- and space-dependent switching} can be observed in Fig.~\ref{fig:dist}, including a higher maximum volume fraction of proliferating cells and a steeper travelling wave front in both ECM and total cell volume fractions.\\

\begin{figure*}[h!]
    \centering
    \includegraphics[scale=1]{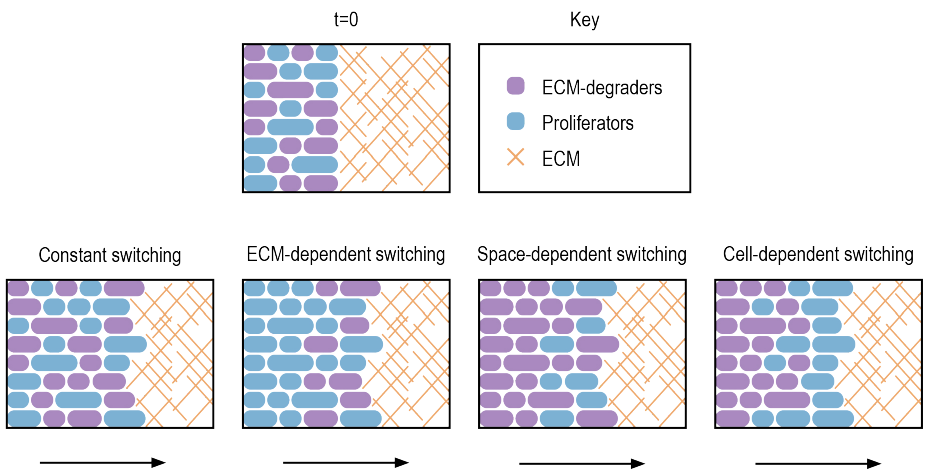}
    \caption{{Schematics demonstrating} the possible cell distributions in the travelling waves observed in this numerical study. Note that there are subtle differences between space- and cell-dependent switching, namely that proliferating cells lead the wave of invasion in isolation during cell-dependent switching.}
    \label{fig:dist_sketch}
\end{figure*}

Overall, it is clear that {constant speed} travelling wave solutions can be observed for the system~\eqref{intro_u1_ND}-\eqref{intro_m_ND} subject to any of the switching functions described in Table~\ref{tab:PStab}. 
Furthermore, the {functional form of} phenotypic switching mechanism {chosen} influences the distribution of cell phenotypic states within the travelling wave{, as {schematised} in Fig.~\ref{fig:dist_sketch}.}

%%%%%params
\subsection{How do the model parameters impact cell migration?}
We now analyse how variations in the model parameters impact the numerically observed travelling wave speed, and determine whether similar trends are observed in the generalist and specialist cell populations. 
In particular, we will consider: 
\begin{itemize}
    \item manipulations of biological parameters specific to the cells, such as the phenotypic switching rate and ECM degradation rate;
    \item manipulations of the environmental conditions, specifically the ECM volume fraction {ahead of the invading wave}. 
\end{itemize}

\subsubsection{Manipulations of cell parameters}
\begin{figure}[htbp]
    \centering
    \includegraphics[scale=0.7]{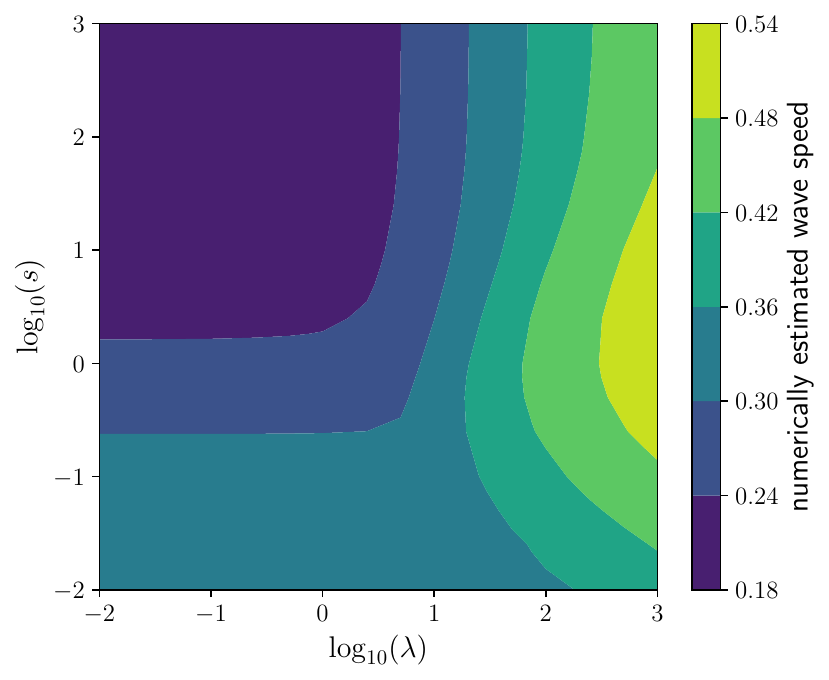}
    \caption{The numerically estimated speed of travelling wave solutions of the system~\eqref{intro_u1_ND}-\eqref{intro_m_ND} subject to the initial conditions for the cells in Eq.~\eqref{IC_up} and {for} the ECM in Eq.~\eqref{IC_m}{,} with cell-dependent phenotypic switching (see Table~\ref{tab:PStab}), {as a function of} the phenotypic switching rate, $s$, and the rescaled ECM degradation rate, $\lambda$. The initial ECM volume fraction ahead of the cells is $m_0=0.5$, the weighting of the specialists towards degrading ECM is $\theta_{S,D}=0.5$ and the width of the {region initially invaded by} migrating cells is $\alpha=1$. For more information regarding the numerical methods used see \ref{app_methods}. }
    \label{fig:cvs_DD}
\end{figure}

\begin{figure*}[htbp]
    \centering
    \includegraphics[scale=0.5]{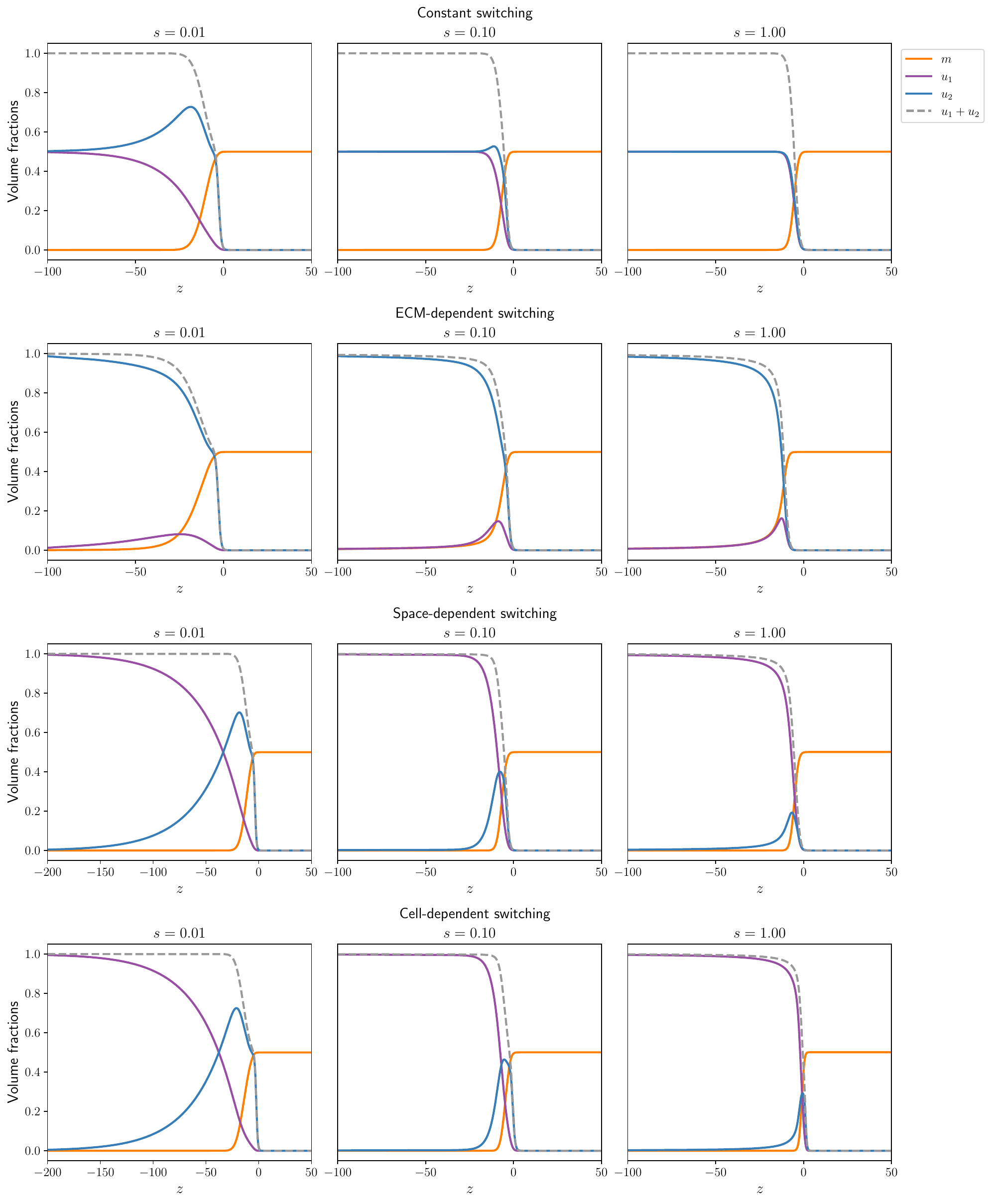}
    \caption{Travelling wave profiles of the solutions of the system~\eqref{intro_u1_ND}-\eqref{intro_m_ND} subject to the initial conditions for the cells in Eq.~\eqref{IC_up} and {for} the ECM in Eq.~\eqref{IC_m}{,} plotted {as a function of} the travelling wave variable $z=x-ct$, where $c$ is the numerically observed {constant} travelling wave speed{. These solutions show} that changing the switching rate changes the distribution of the cell phenotypes within the {invading} wave front. Here, the initial ECM volume fraction ahead of the cells is $m_0=0.5$, the ECM degradation rate is $\lambda=1$, the weighting of the specialists towards degrading ECM is $\theta_{S,D}=0.5$ and the width of {the region initially invaded by} migrating cells is $\alpha=1$ in all cases.  For more information regarding the numerical methods used see \ref{app_methods}. }
    \label{fig:varys}
\end{figure*}

\vspace{2mm}
\paragraph{Variations in the phenotypic switching rate}
{Numerical simulations suggest that v}ariations in the switching rate {generally have} {a small} impact on the cell migration speed when we employ constant, ECM- or space-dependent switching (for further details, the reader is directed to the Supplementary Material~\ref{supp:s12}{, where {phenotypic} switching at different rates in either direction is also considered}).
However, when considering cell-dependent switching we find that changing the switching rate has a significant impact on the speed {of invasion}. 
Fig.~\ref{fig:cvs_DD} reveals that for low values of the ECM degradation rate, $\lambda$, the travelling wave speed increases as the switching rate, $s$, decreases and the wave front becomes smoother (see Figure~\ref{fig:varys}).
However, for {sufficiently} large $\lambda$, ECM degradation dominates over phenotypic switching to determine the travelling wave speed, and {the} maximum {invasion} speed is reached when $s=1$. 
{The optimal switching rate{, in terms of the} fastest speed of invasion{,} is found analytically} in Supplementary Material~\ref{supp:max}.

When considering fast phenotypic switching, following ideas in \cite{crossley2023travelling}, we can {formally} find {expressions} for the travelling wave speed in asymptotic regimes of the ECM degradation rate{,} $\lambda\to0^{+}$ and $\lambda\to\infty${,} that match the {wave} speed observed {in} {numerical} solutions to Eqs.~\eqref{intro_u1_ND}-\eqref{intro_m_ND} subject to the initial conditions for the cells in Eq.~\eqref{IC_up} and {for} the ECM in Eq.~\eqref{IC_m} (see Supplementary Material~\ref{supp:anasy}). 

Despite the phenotypic switching rate, $s$, having little quantitative {impact} on the travelling wave speed for most {moderate} parameter {values} across most phenotypic switching mechanisms considered, {changing $s$ does significantly impact} the distribution of cells within the travelling wave front in all cases.
By looking at the top row of Fig.~\ref{fig:varys}, when we have constant phenotypic switching, we see that increasing the switching rate balances the proportion of cells throughout the wave{, which is {consistent with} analytical {results detailed} in Supplementary Material~\ref{supp:anasy}}. 
As the phenotypic switching rate decreases, {however,} there is a larger proportion of cells in phenotypic state 2 at the front of the wave, and a wider travelling wave profile. 
For ECM-dependent switching, increasing the switching rate increases the proportion of cells in phenotypic state 1 at the wave front, and concentrates them to the front, leading to sharper travelling wave profiles. 
Alternatively, for space- and cell-dependent phenotypic switching mechanisms, increasing the switching rate reduces the volume fraction of cells in phenotypic state 2. 
This reduction is larger with space-dependent phenotypic switching.
Qualitatively, the travelling wave profiles for space- and cell-dependent switching are almost identical (see the bottom two rows of Fig.~\ref{fig:varys}), and increasing the switching rate decreases the maximum volume fraction of cells in phenotypic state 2 at the front of the wave, shortening the tail of the travelling pulse and leading to sharper travelling wave profiles in the total volume fraction of cells. 

\paragraph{Variations in ECM degradation rate}
In the model of a homogeneous cell type invading into the ECM (defined by Eqs.~\eqref{gen_weight_u}-\eqref{gen_weight_m}), the shape and speed of the travelling wave changes as the ECM degradation rate varies \cite{crossley2023travelling}.
Relationships between the rescaled ECM degradation rate in asymptotic regions and travelling wave speed were established previously for the fully non-dimensional model, without weightings, {and} {it was} shown {that} $c\to2^{-}$ as $\lambda\to\infty$ and $c\to2(1-m_0)$ as $\lambda\to0^{+}$ \cite{crossley2023travelling}.

\begin{figure*}[htbp]
    \centering
    \includegraphics[scale=0.55]{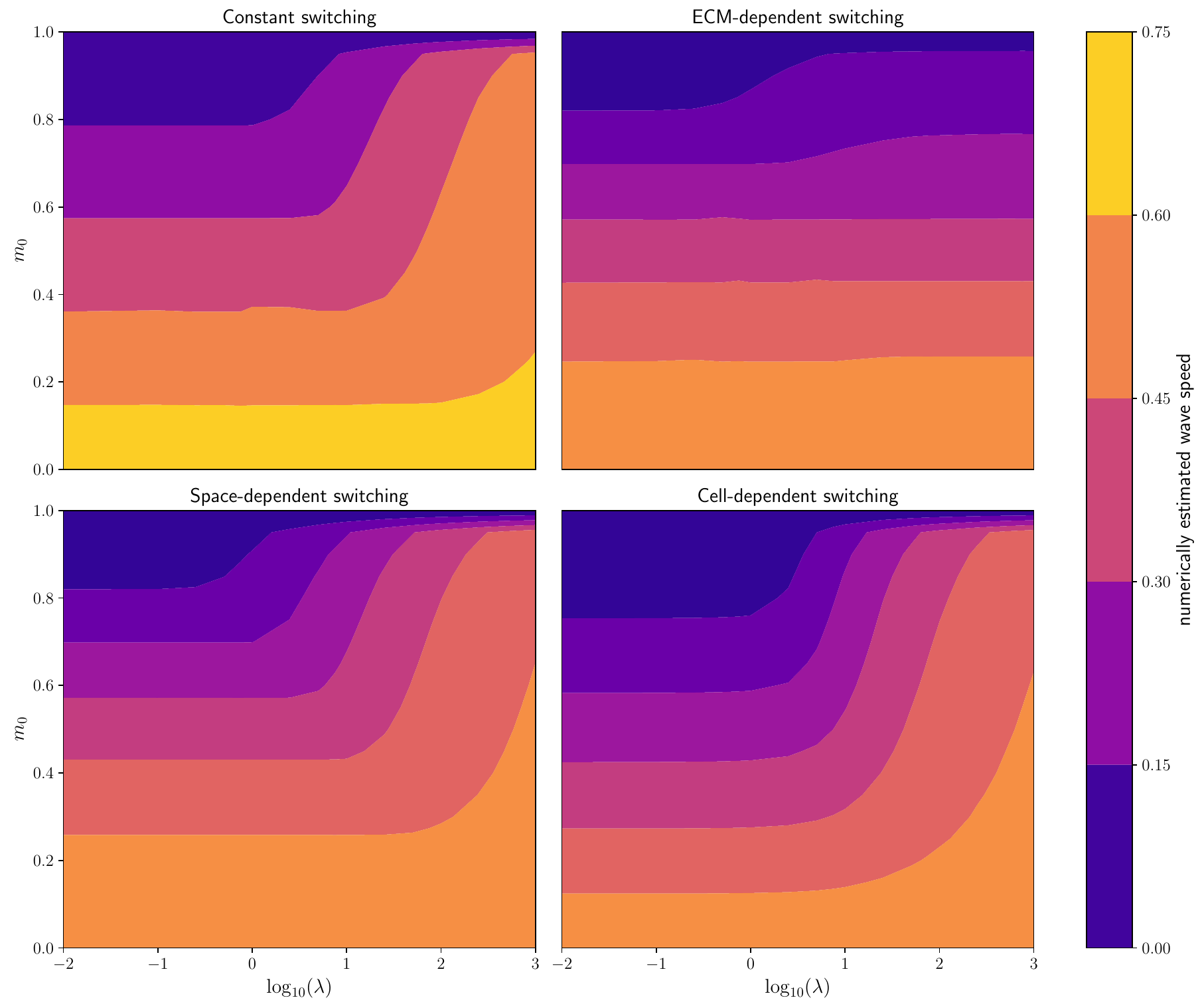}
    \caption{Plots summarising the relationship between the numerically estimated speed of travelling wave solutions of Eqs.~\eqref{intro_u1_ND}-\eqref{intro_m_ND} subject to the initial conditions for the cells in Eq.~\eqref{IC_up} and {for} the ECM in Eq.~\eqref{IC_m}, the initial volume fraction of ECM ahead of the cells, $m_0$, and ECM degradation rate, $\lambda$. 
    Here, the weighting of the specialists towards degrading ECM is $\theta_{S,D}=0.5$, the switching rate is $s=1$ and the width of the region initially invaded by migrating cells is $\alpha=1$ in all cases.  For more information regarding the numerical methods used see \ref{app_methods}.}
    \label{fig:m0_dm_contours}
\end{figure*}

By inspecting Fig.~\ref{fig:m0_dm_contours}, we can see that{,} across all the switching functions that we consider in this work, an increase in the ECM degradation rate {leads to} an increase in the numerically estimated travelling wave speed {when} ECM degradation rates {are} above a critical value.
{We also see} that as $\lambda\to\infty$ convergence in the travelling wave speed {to a constant value} is observed for constant, space-dependent and cell-dependent switching mechanisms, but to different values.
{In Supplementary Material~\ref{supp:anasy}, we perform formal analysis of the system~\eqref{intro_u1_ND}-\eqref{intro_m_ND} for general switching rates that can differ in either direction.
In the particular case where the switching rate is the same in either direction,} we show analytically that, in the fast phenotypic switching regime, $c\rightarrow(1-m_0)\sqrt{1-\theta_{S,D}}$ as $\lambda\to0^{+}$ and $c\rightarrow\sqrt{1-\theta_{S,D}}$ as $\lambda\to\infty$ for constant phenotypic switching. 
Convergence of the numerically estimated travelling wave speed to these values can be seen in Figure~\ref{fig:fastPS_c} of Supplementary Material~\ref{supp:anasy}.
In contrast to {the other switching functions,} ECM-dependent switching is far less sensitive to changes in ECM degradation rates at low initial ECM volume fractions ahead of the cells, and convergence of the travelling wave speed {to a constant value} is not observed within the parameter ranges considered in this work.

From a biological perspective, {the limit} {$\lambda\to\infty$} {is relevant in} describing cells in an aggressive tumour {that have} a very high ability to degrade ECM{,} {which {may} enable them} to invade much faster. 
Alternatively, for a sufficiently small product $\theta_{S,D}\lambda$, specialist cells in phenotypic state 1 should focus more of their ability on movement (i.e., decrease $\theta_{S,D}$) in order to increase {migration} speed, since a small change in the ECM degradation rate alone, when below some threshold value, will minimally impact the migration speed.

\subsubsection{Manipulations of the environmental conditions}
For a generalist cell population, it can be shown analytically and numerically that the speed of the travelling wave of migrating cells increases as {the} initial ECM volume fraction ahead of the cells decreases \cite{crossley2023travelling}. 
Specifically, for small ECM degradation rates (i.e., $\lambda\to0^{+}$), there is a linear relationship between the travelling wave speed, $$c_{{G}}=2(1-m_0)\sqrt{\theta_{G,P}(1-\theta_{G,D}-\theta_{G,P})},$$ and the initial volume fraction of ECM ahead of the wave, $m_0\in[0,1).$

Examining Fig.~\ref{fig:m0_dm_contours}, it is clear that{,} across all four phenotypic switching functions considered, {the speed of the travelling wave of the specialist cell population increases as the initial ECM volume fraction ahead of the cells} decreases.
Biologically, a lower ECM volume fraction corresponds to a less densely packed region of ECM ahead of the cells {which} facilitates faster cell {invasion}.

%%%%%%%%%%%%% Discussion
\section{Discussion}\label{discuss}
Recently, population heterogeneity, such as leaders and followers, has been recognised as an {important} driver of collective cell migration {and has attracted significant} attention \cite{vilchez2021decoding}. 
Here, we have extended a model for a homogeneous generalist population of cells migrating into {the} ECM to explicitly incorporate phenotypic heterogeneity under the migration-proliferation dichotomy. 
We considered {how} distinct phenotypic switching mechanisms {impact} population structure, and {the} dependence {of the travelling wave speed} on {different biological} parameters. 
Specifically, we considered constant switching, ECM-dependent switching, space-dependent switching and cell-dependent switching. 

Initially, we compared {a} homogeneous cell population to a heterogeneous cell population without phenotypic switching. In this case, it {is} clear that a specialist cell population without the ability to change its phenotypic state {can} never {outcompete} a generalist cell population {as} {the model does not admit} travelling wave {solutions}. 
Conversely, {analysis of} the model for a specialist cell population invading into {the} ECM that include{s} phenotypic switching {show{s}} that a heterogeneous cell population can produce travelling waves of invasion with a faster speed than homogeneous generalist populations {that weight their ability towards movement}. 
Moreover, we confirmed that the travelling wave speed {in} the specialist model, irrespective of the phenotypic switching mechanism, depends qualitatively on the ECM degradation rate and initial ECM volume fraction ahead of the cells in the same manner as {for} {the generalist model}.

This work considers phenotypic switching {that is equal in either direction}, {and in this case} the travelling wave speed {is} shown to be independent of the switching rate for constant, ECM- and space-dependent switching. 
{When asymmetric switching rates are considered, {both switching rates impact the speed and distribution of phenotypes in the invading cell population} (see Supplementary Material~\ref{supp:s12}~and~\ref{supp:anasy}).
In the case of cell-dependent switching, increasing the switching rate decreases the travelling wave speed.

Biologically, there exist a number of {factors that could drive} phenotypic {switching}. 
{For example,} direct contact between cell surfaces, or contact on the cell surface from molecules released by neighbouring cells or ECM, {which} provide information about surrounding cell and ECM volume fractions, might cause phenotypic change \cite{tosh2002cells, quesenberry2010cellular}. 
This work demonstrate{s} that the mechanism determining the form of phenotypic switching function has a profound impact on the phenotypic structure of the invading cell population.  
When there is no environmental dependence, a {well-}mixed population of cells {invades} into {the} ECM, with the ratio of cells in phenotypic state 1 to phenotypic state 2 in the bulk being determined by the ratio between the switching rates. 
In ECM-dependent switching {models}, degrading cells in phenotypic state 1 occupy the migrating front, with proliferating cells in the bulk, {as} observed in many {examples of} leader-follower dynamics \cite{konen2017image} where both sub-populations play an important role in driving {cell invasion}.
Space{-} and cell-dependent switching {models} exhibit the opposite distribution, with proliferating cells leading the way{, as observed} \textit{in vivo} and \textit{in vitro} in melanoma spheroid growth, where proliferative clusters form on the outer edges of a larger bulk cluster consisting of migratory cells \cite{campbell2021cooperation}.   
This could be favourable for the population, {for example, if} there are lower energetic requirements {for} proliferat{ion} in low volume fraction regions.
{A further extension of this work would be to consider a general switching function combining the influence of both available space and ECM, to find critical parameter values that determine the transition between the degraders or the proliferators leading the invasive population.}

{The speed of invasion of a cell population alone does not necessarily allow us to distinguish the mechanisms governing collective cell migration. 
{In the case of space- and cell-dependent switching, changing the phenotypic switching rate and observing the changes in the resulting travelling wave speed may indeed be sufficient to distinguish between the two switching mechanisms, where differences in the travelling wave profile are otherwise very subtle.}
However, {in other cases,} the {spatial} structure of each cell sub-population within an invading wave might provide further insights.}
For instance, examination of a detailed population profile from a tissue biopsy in the direction of migration could be used as a predictive tool to distinguish the mechanism underlying cell phenotypic switching which could{,} in turn{,} be used to help develop therapeutic treatments.
Moreover, if such a biopsy revealed {cells in one phenotypic state only} at the front {of the invading wave}, this may indicate that {the rate of} phenotypic switching in one direction far exceed{s the rate of phenotypic} switching in the opposite direction. 
For example, in the case of ECM-dependent switching, if a population of primarily ECM-degrading cells is observed at the migrating front, this suggests that phenotypic switching from proliferative to degrading phenotypic state is {much} faster than from degrading to proliferative.
{Simulations in Supplementary Material~\ref{supp:s12}} simultaneously revealed a trade-off between population structuring and the travelling wave speed, such that when a single population leads the migrating front, the invasion speed is reduced.  
This suggests {that both the spatial profile and speed of invasion are required to distinguish the underlying phenotypic switching mechanisms and that} further possible therapeutic treatments {could be developed to slow tumour growth by preventing switching in one direction, or to speed up wound healing or developmental processes by initiating symmetrical switching.}

There are various possible extensions to this work. 
The biological applicability could be expanded by varying the underlying assumptions of the model and including other possible factors influencing collective cell migration, such as haptotaxis or chemotaxis.
By tailoring this model to a particular biological application, future work could also explore validating the model predictions, estimating model parameters, and establishing the specific form of the phenotypic switching functions by examining appropriate histology data.
Furthermore, the current study looks exclusively at these populations in one spatial dimension, and a future avenue for exploration could be to} extend the model to two or more dimensions. 
In higher dimensions it would be interesting to investigate the
stability {of the {invading} front}  or the presence of spatial structure, such as  ``fingering''{,} which has previously been observed in models of tumour growth containing {sub-populations} with different mobility \cite{lorenzi2016interfaces, drasdo2012modeling} and for tumour growth into heterogeneous ECM \cite{anderson2006tumor}.

In the specialist {population} model, it is unclear how to compare the models with different switching functions since, in the case of ECM-, space- and cell-dependent switching, the switching rate, $s$, is rescaled by a term bounded in $[0,1]$ varying in time. 
This suggests a further extension of this work to include energetic costs of proliferation, movement, ECM degradation and phenotypic switching between states that allows for a more biologically conclusive investigation of which form of phenotypic switching function generates the {highest} {rate of cell invasion}. 
It would also be of interest to compare these model results to those from the underlying discrete model and to perform analysis to {derive} an explicit expression for the travelling wave speed as a function of all the model parameters, defining the parameter {space wherein} the numerically estimated travelling wave speed {matches that predicted by analysis}. 
Additionally, it appears {that} for all of the phenotypic switching mechanisms considered, the travelling wave speed depends linearly on the initial ECM volume fraction and is independent of $\lambda$ for $\lambda\leq\lambda_{c}$, where $\lambda_{c}$ is a critical value.  
Future work might also seek {to determine} this critical value explicitly using asymptotic or boundary layer analysis, by considering a thin layer of cells around the wavefront at the interface with the ECM, which becomes sharp for large values of $\lambda.$

In conclusion, understanding the phenotypic structure of {invading} cell{ular collectives} is an important objective, attracting significant research over recent years. 
The model presented in this study, whilst clearly {simple}, provides compelling {insights} associated with the speed and structure of heterogeneous cell invasion into {the} ECM under various phenotypic switching mechanisms, and provides a basis for more complex and detailed model development and analysis in the future.

\section*{CReDiT authorship contribution statement}
\textbf{Rebecca M. Crossley:} Conceptualization, Methodology, Software, Formal analysis, Writing - Original Draft, Writing - Review \& Editing, Visualization. 
\textbf{Kevin J. Painter:} Conceptualization, Methodology, Writing - Review \& Editing.
\textbf{Tommaso Lorenzi:} Conceptualization, Methodology, Formal analysis, Writing - Review \& Editing.
\textbf{Philip K. Maini:} Conceptualization, Methodology, Formal analysis, Writing - Review \& Editing.
\textbf{Ruth E. Baker:} Conceptualization, Methodology, Formal analysis, Writing - Review \& Editing.

\section*{Declaration of competing interest}
The authors do not have any competing interests to declare.

\section*{Acknowledgements}
RMC is supported by funding from the Engineering and Physical Sciences Research Council (EP/T517811/1) and the Oxford-Wolfson-Marriott scholarship at Wolfson College, University of Oxford.
KJP is a member of INdAM-GNFM and acknowledges “Miur-Dipartimento di Eccellenza” funding to the Dipartimento
di Scienze, Progetto e Politiche del Territorio (DIST).
TL gratefully acknowledges support from the Italian Ministry of University and Research (MUR) through the grant PRIN 2020 project (No. 2020JLWP23) “Integrated Mathematical Approaches to Socio-Epidemiological Dynamics” (CUP: E15F21005420006) and the grant PRIN2022-PNRR project (No. P2022Z7ZAJ) “A Unitary Mathematical Framework for Modelling Muscular Dystrophies” (CUP: E53D23018070001). 
TL gratefully acknowledges also support from the Istituto Nazionale di Alta Matematica (INdAM) and the Gruppo Nazionale per la Fisica Matematica (GNFM).
{PKM would like to thank the Department of Mathematical Sciences of the Politecnico di Torino for their kind hospitality during his collaborative visit.}
The authors would like to thank the Isaac Newton Institute for Mathematical Sciences, Cambridge, for support and hospitality during the programme Mathematics of Movement where work on this paper was undertaken. This work was supported by EPSRC grant no EP/R014604/1.
{This work was also supported by a grant from the Simons Foundation (MP-SIP-00001828, REB).}

%%%%%%%%%%%%%%%%%%%%%%%%%% APPENDIX 
%\newpage
\begin{appendix}

\setcounter{figure}{0}  

\section{Formal derivation of the continuum model~\eqref{intro_u1_ND}-\eqref{intro_m_ND} from an underlying individual-based model} \label{model_deriv_app}
We begin by developing a simple one-dimensional, on-lattice, individual-based model of {two} distinct cell phenotypes invading into the ECM, in the presence of volume-filling effects, where motility and proliferation are reduced in higher density regions of space. 
Volume-filling effects warrant being accounted for in this way since research shows that cells and ECM both regulate cell movement and proliferation. 
It was experimentally demonstrated that these effects are indeed inhibitory in high density regions, and stimulatory in low ECM regions \cite{dickinson1993stochastic, nusgens1984collagen, weinberg1985regulation, yoshizato1985growth}.
This model derivation follows directly from the ideas in~\cite{crossley2023travelling} extended to multiple {cell sub-}populations, and uses {mean-field} assumption ideas \cite{penington2011building} from multi-species exclusion processes \cite{simpson2009multi}.
Following the go-or-grow assumption studied in \cite{hatzikirou2012go, stepien2018traveling}, we consider a phenotypic {trade-off} between cells {that are degrading the ECM and migrating, and those that are proliferating}, where those that proliferate {do not move}.
We then coarse-grain this model to formally derive a corresponding PDE model that comprises a system of coupled differential equations for the densities of cells and ECM. 

\subsection{Individual-based model}
In this model, cells are represented as discrete individuals with finite volume. 
We consider cells with the ability to change phenotypic state that move randomly on a one-dimensional uniform lattice, which constitutes the spatial domain, proliferate, and degrade the surrounding ECM, which is composed of discrete elements of the same volume as the cells.

Let the number of cells of phenotype $i=\{1,2\}$ and ECM elements in lattice site $j=1,\dots,\,J$ of width $\Delta$, at time $\tilde{t}\in\mathbb{R}_{+}$ of realisation $r=1,\dots, R$ {of the model} be denoted by $u_{i,j}^r(\tilde{t})$ and $m_j^{r}(\tilde{t})$, respectively.

\paragraph{Occupancy level of the lattice sites}
In order to incorporate volume-filling effects into the model, we prescribe that each lattice site has a maximum total occupancy level of $N$ cells and ECM elements, so that
\begin{equation}
    0\leq \sum_{i=1}^{2}u_{i,j}^{r}(\tilde{t})+m_{j}^{r}(\tilde{t})\leq N.\label{Occ_level}\nonumber
\end{equation}

\paragraph{Probability of cell movement}
A cell with phenotype $i$ will attempt a movement event in a time step $\tau$ with probability $p_{i,\text{m}}\in[0,1]$, whereby the attempted movement from lattice site $j$ to either of the neighbouring lattice sites $j\pm 1$ occurs with equal probability $1/2$. 
We assume that the probability of a successful move decreases linearly with the occupancy level of the target site, such that it is zero when the target site is full, and one when it is empty. 
{Hence}, we define the probability of a movement to the left, $T_{j-}^{i,{\rm{m}}^{r}}(\tilde{t})$, or right, $T_{j+}^{i,{\rm m}^{r}}(\tilde{t})$, in $[\tilde{t}, \tilde{t}+\tau)$ of realisation $r$ as \begin{equation}
 T_{j\pm}^{i,{\rm m}^{r}}(\tilde{t})=\frac{p_{i,\rm m}}{2}\bigg(1-\frac{\sum_{i=1}^{2}u_{i,j\pm1}^{r}(\tilde{t}) +m_{j\pm1}^{r}(\tilde{t})}{N}\bigg). \label{total_prob_move}\nonumber
\end{equation}
Zero flux boundary conditions are implemented such that any attempted move outside of the {spatial} domain {is} aborted.

\paragraph{Probability of cell proliferation}
A cell with phenotype $i$ in lattice site $j$ attempts a proliferation event, placing a daughter cell of equal size into the same lattice site, during time step $\tau$ with probability $p_{i,\rm p}\in[0,1].$ 
We assume the probability of a successful proliferation event, where one cell is replaced by two daughter cells with the same heritable phenotypic state as the parent cell, decreases linearly with the occupancy level of the lattice site, such that the probability of a successful proliferation event, $T_{j}^{i,{\rm p}^{r}}(\tilde{t})$, in time interval $[\tilde{t}, \tilde{t}+\tau)$ of realisation $r$ {of the model is} 
\begin{equation}
    T_{j}^{i,{\rm p}^{r}}(\tilde{t}) = p_{i,\rm p} \bigg(1-\frac{\sum_{i=1}^{2}u_{i,j}^{r}(\tilde{t})+m_{j}^{r}(\tilde{t})}{N}\bigg).
    \label{total_prob_prolif}\nonumber
\end{equation}

\paragraph{Probability of cell phenotype change}
{During time interval $[\tilde{t}, \tilde{t}+\tau)$ of realisation $r$ of the model, a  cell with phenotype $i$ will {change to} phenotype $l$ with probability  
\begin{equation}
f_{i\to l}\big(u_{1,j}^r(\tilde{t}),\,u_{2,j}^r(\tilde{t}),\, m_j^r(\tilde{t})\big),\nonumber
\end{equation}
where $f_{i\to l}: \mathbb{R}_{+}^{3}\rightarrow[0,1].$}

\paragraph{Probability of ECM degradation}
During the time interval $[\tilde{t}, \tilde{t}+\tau)$ of realisation $r$, an element of ECM in lattice site $j$ is degraded by cells with phenotype $i$ in the same lattice site with a probability $p_{i,\rm d}\in[0,1].$ Therefore the overall degradation {rate} per unit element of ECM by cells with phenotype $i$, $T_{j}^{i, {\rm d}^{r}}(\tilde{t}),$ is 
\begin{equation}
    T_{j}^{i, {\rm d}^{r}}(\tilde{t})= p_{i,\rm d}u_{i,j}^{r}(\tilde{t}).\nonumber
\end{equation}

\subsection{Coarse-grained model}
To derive a coarse-grained description of the individual-based model, we introduce the average occupancy of each lattice site $j$ by cells of type $i=\{1,\,2\}$ and ECM at time $\tilde{t}$ over $R$ total realisations of the model as
\begin{equation}
   \langle u_{i,j}(\tilde{t}) \rangle = \frac{1}{R} \sum_{r=1}^R u^r_{i,j}(\tilde{t}) \;\; \text{and} \;\; \langle m_j(\tilde{t}) \rangle= \frac{1}{R} \sum_{r=1}^R m^r_j(\tilde{t}). \nonumber
\end{equation}

\paragraph{Coarse-grained model of cell dynamics}
We write a conservation equation using mean-field approximations and independence of lattice sites by considering changes in the average occupancy in the lattice site $j$ during the time interval $[\tilde{t}, \tilde{t}+\tau)$ to give:
 \begin{align} \label{CE_u}
    \langle u_{i,j}(\tilde{t}+\tau) \rangle = \langle u_{i,j}(\tilde{t}) \rangle &+ \frac{p_{i, \rm m}}{2}\langle u_{i,j+1}(\tilde{t}) \rangle \bigg(1- \frac{\sum_{i=1}^{2}\langle u_{i,j}(\tilde{t})\rangle + \langle m_{j}(\tilde{t})\rangle}{N}\bigg)\nonumber\\&+\frac{p_{i, \rm m}}{2}\langle u_{i,j-1}(\tilde{t}) \rangle \bigg(1- \frac{\sum_{i=1}^{2}\langle u_{i,j}(\tilde{t})\rangle + \langle m_{j}(\tilde{t})\rangle}{N}\bigg) \nonumber\\&-\frac{p_{i,\rm m}}{2}\langle u_{i,j}(\tilde{t}) \rangle \bigg(1- \frac{\sum_{i=1}^{2}\langle u_{i,j+1}(\tilde{t})\rangle + \langle m_{j+1}(\tilde{t})\rangle}{N}\bigg) \nonumber\\&-\frac{p_{i,\rm m}}{2}\langle u_{i,j}(\tilde{t}) \rangle \bigg(1- \frac{\sum_{i=1}^{2}\langle u_{i,j-1}(\tilde{t})\rangle + \langle m_{j-1}(\tilde{t})\rangle}{N}\bigg) \nonumber\\ &+ p_{i,\rm p}\langle u_{i,j}(\tilde{t}) \rangle \bigg(1- \frac{\sum_{i=1}^{2}\langle u_{i,j}(\tilde{t})\rangle + \langle m_{j}(\tilde{t})\rangle}{N}\bigg) \nonumber\\ &  - f_{i\to l}\big(\langle u_{1,j}(\tilde{t})\rangle,\,\langle u_{2,j}(\tilde{t})\rangle,\, \langle m_j(\tilde{t})\rangle\big) \langle u_{i, j}(\tilde{t}) \rangle \nonumber\\&+ f_{l\to i}\big(\langle u_{1,j}(\tilde{t})\rangle,\,\langle u_{2,j}(\tilde{t})\rangle,\,\langle m_j(\tilde{t})\rangle\big) \langle u_{l, j}(\tilde{t}) \rangle.
\end{align}
Rearranging Eq. \eqref{CE_u}, and dividing by $\tau$, we {find} 
\begin{align}
    &\frac{\langle u_{i,j}(\tilde{t}+\tau) \rangle - \langle u_{i,j}(\tilde{t}) \rangle }{\tau}\nonumber\\ 
    &= \frac{p_{i, \rm m}\Delta^2}{2\tau}\bigg(1- \frac{\sum_{i=1}^{2}\langle u_{i,j}(\tilde{t})\rangle + \langle m_{j}(\tilde{t})\rangle}{N}\bigg) \Bigg[\frac{\langle u_{i,j+1}(\tilde{t}) \rangle -2 \langle u_{i,j}(\tilde{t}) \rangle+ \langle u_{i,j-1}(\tilde{t}) \rangle}{\Delta^2}\Bigg] \nonumber \\ &\quad+ \frac{p_{i, \rm m}\Delta^2}{2\tau} \langle u_{i,j}(\tilde{t})\rangle  \Bigg[\frac{(\sum_{i=1}^{2}\langle u_{i,j+1}(\tilde{t})\rangle + \langle m_{j+1}(\tilde{t})\rangle)-2(\sum_{i=1}^{2}\langle u_{i,j}(\tilde{t})\rangle + \langle m_{j}(\tilde{t})\rangle)+(\sum_{i=1}^{2}\langle u_{i,j-1}(\tilde{t})\rangle + \langle m_{j-1}(\tilde{t})\rangle)}{\Delta^2}\Bigg]\nonumber \\ &\quad+ \frac{p_{i,\rm p}}{\tau}\langle u_{i,j}(\tilde{t}) \rangle \bigg(1- \frac{\sum_{i=1}^{2}\langle u_{i,j}(\tilde{t})\rangle + \langle m_{j}(\tilde{t})\rangle}{N}\bigg)-\frac{1}{\tau}f_{i\to l}\big(\langle u_{1,j}(\tilde{t})\rangle,\,\langle u_{2,j}(\tilde{t})\rangle,\, \langle m_j(\tilde{t})\rangle\big) \langle u_{i, j}(\tilde{t}) \rangle \nonumber\\&\ \quad + \frac{1}{\tau}f_{l\to i}\big(\langle u_{1,j}(\tilde{t})\rangle,\,\langle u_{2,j}(\tilde{t})\rangle,\,\langle m_j(\tilde{t})\rangle\big) \langle u_{l, j}(\tilde{t}) \rangle. \label{above}
\end{align}
Dividing {Eq.~\eqref{above}} by $\Delta$ and Taylor expanding both sides, before taking limits as $\Delta,\,\tau\to0$, we obtain a description for cell density dynamics in terms of variables $\tilde{u}_i(\tilde{x},\tilde{t})$ and $\tilde{m}(\tilde{x}, \tilde{t})$, which are the continuum equivalents of the number density of cells, $\langle u_{i,j}(\tilde{t})\rangle/\Delta$, and the density of ECM, $\langle m_{j}(\tilde{t})\rangle/(\tilde{\mu}\Delta)$, at $\tilde{x}\in\mathbb{R},\, \tilde{t}\in\mathbb{R}_{+},$ {respectively,} where $\tilde{\mu}$ represents the number of cells equivalent to a unit mass of ECM and {serves} as a conversion factor between the density of ECM, as defined by mass of ECM per unit volume, and the number density of ECM elements, given by $\tilde{\mu}\tilde{m}(\tilde{x},\tilde{t})$.
{Under} the following {scalings}: 
\begin{align}
     &\lim_{\Delta\to0}\frac{N}{\Delta}=\tilde{K}, \qquad \lim_{\tau\to0}\frac{p_{i,\rm p}}{\tau} = \tilde{r_i}, \qquad \lim_{\Delta, \tau \to 0}\frac{p_{i,\rm m}\Delta^2}{\tau} = \tilde{D_i}, \qquad \nonumber\\ 
     &\lim_{\tau\to0} \frac{1}{\tau} f_{i\to l}\big(\langle u_{1,j}(\tilde{t})\rangle,\,\langle u_{2,j}(\tilde{t})\rangle,\, \langle m_j(\tilde{t})\rangle\big) = \tilde{\gamma_{il}}(\tilde{u_1}, \, \tilde{u_2}, \, \tilde{m}), \nonumber
\end{align}
{for all $i,\,l=\{1,2\}$, $i\neq l,$ we find}
\begin{align}
    \frac{\partial \tilde{u_i}}{\partial \tilde{t}} &= \tilde{D_i}\Bigg[ \bigg(1-\frac{\sum_{i=1}^{2} \tilde{u}_i+\tilde{\mu}\tilde{m}}{\tilde{K}}\bigg)\frac{\partial^2\tilde{u}_i}{\partial \tilde{x}^2}+\tilde{u}_i \frac{\partial ^2}{\partial \tilde{x}^2}\bigg(\frac{\sum_{i=1}^{2} \tilde{u}_i+\tilde{\mu}\tilde{m}}{\tilde{K}}\bigg)\Bigg]+\tilde{r_i}\tilde{u}_i\bigg(1-\frac{\sum_{i=1}^{2} \tilde{u}_i+\tilde{\mu}\tilde{m}}{\tilde{K}}\bigg)\nonumber \\ &\qquad - \tilde{\gamma_{il}}(\tilde{u}_1,\, \tilde{u}_2,\,\tilde{m}) \tilde{u}_i +  \tilde{\gamma_{li}}(\tilde{u}_1,\, \tilde{u}_2,\,\tilde{m}) \tilde{u}_l, \nonumber \\ 
    &= \tilde{D_i} \frac{\partial}{\partial \tilde{x}}\Bigg[ \bigg(1-\frac{\sum_{i=1}^{2} \tilde{u}_i+\tilde{\mu}\tilde{m}}{\tilde{K}}\bigg)\frac{\partial \tilde{u}_i}{\partial \tilde{x}}+\tilde{u}_i \frac{\partial }{\partial \tilde{x}}\bigg(\frac{\sum_{i=1}^{2} \tilde{u}_i+\tilde{\mu}\tilde{m}}{\tilde{K}}\bigg)\Bigg]+\tilde{r_i}\tilde{u}_i\bigg(1-\frac{\sum_{i=1}^2 \tilde{u}_i+\tilde{\mu}\tilde{m}}{\tilde{K}}\bigg)\nonumber \\ &\qquad - \tilde{\gamma_{il}}(\tilde{u}_1,\, \tilde{u}_2,\,\tilde{m}) \tilde{u}_i + \tilde{\gamma_{li}}(\tilde{u}_1,\, \tilde{u}_2,\,\tilde{m}) \tilde{u}_l,\nonumber
\end{align}
where $\tilde{x}\in\mathbb{R}$ and $\tilde{t}\in\mathbb{R}_{+}.$

\paragraph{Coarse-grained model of ECM dynamics}
{In the same way,} using probabilistic assumptions of mean-field type, we can write the following conservation equation for the evolution of ECM elements in a lattice site $j$ over the time interval $[\tilde{t}, \tilde{t}+\tau)$:
\begin{equation} \label{CE_m}
    \langle m_{j}(\tilde{t}+\tau)\rangle = \langle m_{j}(\tilde{t})\rangle -\sum_{i=1}^{2} p_{i,\rm d} \langle u_{i,j}(\tilde{t})\rangle\langle m_{j}(\tilde{t})\rangle.
\end{equation}
By rearranging Eq.~\eqref{CE_m}, dividing by $\Delta$ and $\tau$ and taking limits as $\Delta,\,\tau\to0$ under the scaling 
\begin{equation}
    \tilde{\lambda_i}=\lim_{\Delta,\tau\to0}\frac{\Delta p_{i,\rm d}}{\tau},\nonumber
\end{equation}
we obtain the following differential equation governing the dynamics of ECM density $\tilde{m}(\tilde{x},\tilde{t})$ over time: 
\begin{equation}
    \frac{\partial \tilde{m}}{\partial \tilde{t}}=-\sum_{i=1}^{2}\tilde{\lambda_i} \tilde{m}\tilde{u_i},\nonumber
\end{equation}
where $\tilde{x}\in\mathbb{R}$ and $\tilde{t}\in\mathbb{R}_{+}.$
The parameter $\tilde{\lambda_i}\geq0$ describes the degradation rate of ECM per cell in phenotypic state $i$.

\paragraph{Full system of equations}
For phenotypes $i=\{1,\,2\}$, $i\neq l$, the full system of equations at population level is given by:
\begin{align}
\frac{\partial \tilde{u_i}}{\partial \tilde{t}} &= \tilde{D_i} \frac{\partial}{\partial \tilde{x}}\Bigg[ \bigg(1-\frac{\sum_{i=1}^{2} \tilde{u}_i+\tilde{\mu}\tilde{m}}{\tilde{K}}\bigg)\frac{\partial \tilde{u}_i}{\partial \tilde{x}}+\tilde{u}_i \frac{\partial }{\partial \tilde{x}}\bigg(\frac{\sum_{i=1}^{2} \tilde{u}_i+\tilde{\mu}\tilde{m}}{\tilde{K}}\bigg)\Bigg]\nonumber+\tilde{r_i}\tilde{u}_i\bigg(1-\frac{\sum_{i=1}^{2} \tilde{u}_i+\tilde{\mu}\tilde{m}}{\tilde{K}}\bigg)\nonumber \\ 
&\qquad - \tilde{\gamma_{il}}(\tilde{u}_1,\, \tilde{u}_2,\,\tilde{m}) \tilde{u}_i +{\tilde{\gamma_{li}}}(\tilde{u}_1,\, \tilde{u}_2,\,\tilde{m}) \tilde{u}_l, \nonumber \\
\frac{\partial \tilde{m}}{\partial \tilde{t}}&=-\sum_{i=1}^{2}\tilde{\lambda_i} \tilde{m}\tilde{u_i}. \nonumber
\end{align}

\subsection{Two populations of specialists with volume-filling}
Now consider the case where cells {in phenotypic state} 1 can move and degrade ECM {only}, and cells {in phenotypic state} 2 are only able to proliferate{, following the go-or-grow hypothesis}.
{We} write $\tilde{\lambda_i}$ as the rate of ECM degradation and $\tilde{r_i}$ as the proliferation rate {of} cells in phenotypic state $i$ {and} assume $$\tilde{\lambda_2}=0, \quad \tilde{r_1}=0, \quad \tilde{D}_2=0, \quad \tilde{\lambda_1}\in\mathbb{R}_{+} \quad \text{and} \quad \tilde{r_2}\in\mathbb{R}_{+}.$$

{Without loss of generality, we assume that cells have a total weighting, $T=1$, to distribute across the mechanisms governing the cells' migration}, irrelevant of their phenotypic state. 
As such, we introduce $\theta_{S,D}\in[0,T]$ to describe the weighting of cells in phenotypic state 1 towards degrading ECM, and
$(T-\theta_{S,D})$ describes the remaining weighting for movement. 

Under the aforementioned assumptions, the following system of {differential} equations {describes the evolution of cell and ECM densities over time:}
\begin{align}
     \frac{\partial \tilde{u_1}}{\partial \tilde{t}} &=(1-\theta_{S,D})\tilde{D_1} \frac{\partial}{\partial \tilde{x}}\Bigg[ \bigg(1-\frac{\sum_{i=1}^{2} \tilde{u}_i+\tilde{\mu}\tilde{m}}{K}\bigg)\frac{\partial \tilde{u}_1}{\partial \tilde{x}}+\tilde{u}_1 \frac{\partial }{\partial \tilde{x}}\bigg(\frac{\sum_{i=1}^{2} \tilde{u}_i+\tilde{\mu}\tilde{m}}{K}\bigg)\Bigg] \nonumber \\ &\qquad\qquad+\tilde{u}_2 \tilde{\gamma}_{21}(\tilde{u}_1, \tilde{u}_2, \tilde{m})-\tilde{u}_1 \tilde{\gamma}_{12}(\tilde{u}_1, \tilde{u}_2, \tilde{m}), \label{u1_dim}\\
     \frac{\partial \tilde{u_2}}{\partial \tilde{t}}&= \tilde{r_2}\tilde{u}_2\bigg(1-\frac{\sum_{i=1}^{2} \tilde{u}_i+\tilde{\mu}\tilde{m}}{K}\bigg)\nonumber\\&\qquad\qquad-\tilde{u}_2 \tilde{\gamma}_{21}(\tilde{u}_1, \tilde{u}_2, \tilde{m})+\tilde{u}_1 \tilde{\gamma}_{12}(\tilde{u}_1, \tilde{u}_2, \tilde{m}), \\
    \frac{\partial \tilde{m}}{\partial \tilde{t}}  &=-\theta_{S,D}\tilde{\lambda_1} \tilde{m}\tilde{u_1}, \label{m_dim}
\end{align}
where the diffusion {coefficient} of cells of type 1 is given by $\tilde{D_1}\in\mathbb{R}_{+}$, the switching between cell phenotypic states $p\in\{1,2\}$ is given by the functions  
\begin{equation*}
    \tilde{\gamma}_{12}:\mathbb{R}_{+}^3\rightarrow\mathbb{R}_{+} \qquad \text{and} \qquad \tilde{\gamma}_{21}:\mathbb{R}_{+}^3\rightarrow\mathbb{R}_{+},
\end{equation*}
where $\tilde{\gamma}_{12}$ represents the rate of switching from phenotypic state 1 to 2, and $\tilde{\gamma}_{21}$ represents the rate of switching from phenotypic state 2 to 1, $\theta_{S,D}\in[0,1]$ is the weighting of cells in phenotypic state 1 to degrade ECM, {and} $\tilde{\lambda_1}\in\mathbb{R}$ is the ECM degradation rate by cells in phenotypic state 1 and $\tilde{x}\in\mathbb{R}.$

\section{Non-dimensionalisation of Eqs.~\eqref{u1_dim}-\eqref{m_dim}}\label{NDapp}
Without loss of generality, we introduce the following non-dimensional variables:
{
\begin{equation*}
    x = \sqrt{\dfrac{\tilde{r_2}}{\tilde{D_1}}}\tilde{x}, \quad t = \tilde{r_2}\tilde{t}, \quad u_1=\dfrac{\tilde{u_1}}{K}, \quad u_2=\dfrac{\tilde{u_2}}{K}, \quad m=\dfrac{\mu\tilde{m}}{K} ,
\end{equation*}
alongside the following non-dimensional form of the switching functions
\begin{align*}
    \gamma_{12}=\;\gamma_{12}(u_1, u_2, m)= & \;\dfrac{1}{\tilde{r_2}}\tilde{\gamma_{12}}(\tilde{u_1}, \tilde{u_2}, \tilde{m}),  \nonumber \\ \gamma_{21}=\;\gamma_{21}(u_1, u_2, m)= & \;\dfrac{1}{\tilde{r_2}}\tilde{\gamma_{21}}(\tilde{u_1}, \tilde{u_2}, \tilde{m}),
\end{align*}
which, substituting into Eqs.~\eqref{u1_dim}-\eqref{m_dim}, {yields} the non-dimensional system
\begin{align}
\frac{\partial {u_1}}{\partial {t}} &=(1-\theta_{S,D})\frac{\partial}{\partial {x}}\Bigg[ \big(1-u_1-u_2-m\big)\frac{\partial {u}_1}{\partial {x}}+{u_1} \frac{\partial }{\partial {x}}\big(u_1+u_2+m\big)\Bigg]\nonumber \\&\qquad\qquad+{{u}_2}{\gamma_{21}(u_1, u_2, m)}-{u}_1{\gamma_{12}(u_1, u_2, m)},\nonumber \\
\frac{\partial {u_2}}{\partial {t}}&= {{u}_2}\big(1-u_1-u_2-m\big)\nonumber \\ &\qquad-{{u}_2}{\gamma_{21}(u_1, u_2, m)}+{u}_1{\gamma_{12}(u_1, u_2, m)}, \nonumber\\
\frac{\partial {m}}{\partial {t}}  &=-\theta_{S,D}\lambda {m}{u_1}, \nonumber
\end{align}
}
where we have introduced the non-dimensional parameter 
\begin{equation*}
    \lambda=\frac{\tilde{\lambda_1}K}{\tilde{r_2}},
\end{equation*}
representing the rescaled ECM degradation rate, noting that $\theta_{S,D}\in[0,1]$ is already a dimensionless parameter.

%%%%%%%%%%%%%%%%%%%%%%% SHSS
\section{Spatially homogeneous steady states} \label{app_SHSS}

In this section, we perform a steady state analysis for the system~\eqref{intro_u1_ND}-\eqref{intro_m_ND} subject to each of the phenotypic switching functions listed in Table~\ref{tab:PStab}. 
Since we are interested in travelling wave solutions, we seek spatially homogeneous steady states. 

For all phenotypic switching mechanisms we consider, the spatially homogeneous steady states satisfy
\begin{align}
    u_2(1-u_1-u_2-m)&=0, \label{SHSS2}\\
    m u_1 &=0. \label{SHSS3} 
\end{align}
Eq.~\eqref{SHSS3} implies that either $m=0$, or $u_1=0,$ and Eq~\eqref{SHSS2} gives {$u_2=0$} or $u_1+u_2+m=1.$

\subsection{Constant switching} \label{SHSS_CS}
For the case that $\gamma_{12}(u_1,\,u_2,\,m)=\gamma_{21}(u_1,\,u_2,\,m)=s$, the spatially homogeneous steady states of the system~\eqref{intro_u1_ND}-\eqref{intro_m_ND} also satisfy
\begin{align*}
    -u_1+u_2&=0, 
\end{align*}
{and} the {continuum of} spatially homogeneous steady states {is given by} 
\begin{align*}
    \mathcal{A}_1&\coloneqq (u_1^{*},\, u_2^{*},\, m^{*}) = (0, \,0, \,\bar{m}),  \\ 
    \mathcal{A}_2&\coloneqq (u_1^{*},\, u_2^{*},\, m^{*}) = \bigg(\bar{u_1}, \, \bar{u_2}, \, 0\bigg),
\end{align*}
{where $\bar{u_1},\, \bar{u_2},\, \bar{m}\in[0,1]$, prescribed by initial conditions{, and $\bar{u_1}=\bar{u_2}=0.5$} (see Supplementary Material~\ref{supp:s12} for values of $\bar{u_1}$ and $\bar{u_2}$ when switching rates differ in either direction.}
The steady state $ \mathcal{A}_1$ describes {the case} with no cells present and only ECM. 
Alternatively, $\mathcal{A}_2$ describes a mixed population of cells, and no ECM, where the ratio between cells in phenotypic state 1 and 2 {is} determined by the {phenotypic} switching rates {in either direction (see Supplementary Fig.~\ref{fig:CS_S_ratio}}).

\subsection{ECM-dependent switching} \label{SHSS_LM}
When we consider ECM-dependent switching, we find {that} {the} spatially homogeneous steady states must satisfy Eqs.~\eqref{SHSS2}~and~\eqref{SHSS3} along with 
\begin{align*}
    -u_1(1-m)+u_2m&=0.  
\end{align*}
By the same arguments as before, the resulting steady states are given by 
\begin{align*}
    \mathcal{B}_1&\coloneqq (u_1^{*},\, u_2^{*},\, m^{*}) = (0, \,0, \,\bar{m}),  \\ 
    \mathcal{B}_2&\coloneqq (u_1^{*},\, u_2^{*},\, m^{*}) = (0, \, 1, \, 0). 
\end{align*}
Once again, the steady state described by no cells and only ECM  (far ahead of the travelling wave) is given by $\mathcal{B}_1$, but {now} $\mathcal{B}_2$ describes a steady state consisting only of cells in phenotypic state 2. 

\subsection{Space-dependent switching} \label{SHSS_AS}
By considering phenotypic switching dependent on available space, the spatially homogeneous steady states satisfy
\begin{align*}
    -s u_1(1-u_1-u_2-m)+s u_2(u_1+u_2+m)&=0, 
\end{align*}
such that the spatially homogeneous steady states are given by 
\begin{align}
    \mathcal{C}_1&\coloneqq (u_1^{*},\, u_2^{*},\, m^{*}) = (0, \,0, \,\bar{m}),  \label{1C}\\ 
    \mathcal{C}_2&\coloneqq (u_1^{*},\, u_2^{*},\, m^{*}) = (1, \, 0, \, 0).  \label{2C}
\end{align}
$\mathcal{C}_1$ is the same steady state described for constant and ECM-dependent switching, where only ECM is present.
$\mathcal{C}_2$ represents a steady state with only cells in phenotypic state 1, and no cells in phenotypic state 2 or ECM present.

\subsection{Cell-dependent switching} \label{SHSS_DD}
Finally, the spatially homogeneous steady states under cell-dependent phenotypic switching satisfy Eqs.~\eqref{SHSS2}-\eqref{SHSS3} and 
\begin{align*}
    -s u_1(1-u_1-u_2)+s u_2(u_1+u_2)&=0,
\end{align*}
to give the spatially homogeneous steady states, $\mathcal{C}_1$ and $\mathcal{C}_2$ {(see Eqs.~\eqref{1C}~and~\eqref{2C}), that are the same as those observed under} space-dependent switching.

%%%%%%%%%%%%%%%%%%%%%%%%%%%% Num sims
\section{Numerical simulation methods} \label{app_methods}
The system~\eqref{intro_u1_ND}-\eqref{intro_m_ND} subject to zero flux boundary conditions and initial conditions for the cells as in Eq.~\eqref{IC_up}, and {for} the ECM as in Eq.~\eqref{IC_m} are solved numerically using the method of lines on a {one-dimensional} spatial domain $x\in[0,L]$, where $L>0$ is chosen to be sufficiently large to remove the impacts of the boundary conditions {and enable convergence to a constant speed travelling wave}. 

To employ the method of lines, the spatial domain is uniformly discretised into $Q$ spatial points, with separation $h$. 
An explicit central differencing scheme, as described in \cite{strobl2020mix}, is then employed to solve the system, {taking} the following form:
\begin{align}
    \frac{\partial}{\partial x}\bigg[D\frac{\partial a}{\partial x}\bigg]&\approx \frac{1}{2h^2}\bigg[(D_{q-1}+D_{q})a_{q-1}+(D_i+D_{q+1})a_{q+1}-(D_{q-1}+2D_{q}+D_{q+1})a_q\bigg],\nonumber
\end{align}
where $a_{q}$ represents the value of {the function} $a$ at the spatial point $x_q.$ 
The system~\eqref{intro_u1_ND}-\eqref{intro_m_ND} can {then} be re-written as {a} system of $3Q$ ordinary differential equations, which is solved with zero flux boundary conditions by simulating the ghost points $x_{-1}$ and $x_{Q+1}$ outside of the initial spatial domain, as described in~\cite{morton2005numerical}. 
The remaining system of equations, which has been discretised in space, is then solved numerically in python using the built-in solver {
\tt{scipy.integrate.solve\_ivp}} with the explicit Runge-Kutta integration method of order 5 and time step $\Delta t=0.1$.

To {estimate} the wave speeds {numerically}, for each time point that we save a solution, we interpolate to find $X(t)$ such that {$$u_1(X(t),t)+u_2(X(t),t)=y^{*}\in(0,1),$$ where we choose $y^{*}=0.1$ arbitrarily} and then calculate 
\begin{equation}
    c_{\text{estimated}}(t,t+\Delta t) = \dfrac{X(t+\Delta t)-X(t)}{\Delta t}.\nonumber
\end{equation}
When the calculated wave speeds are observed to have converged to a constant speed, such that the difference between two subsequent measurements is of an order smaller than the order of error of the numerical scheme, we record this as the travelling wave speed estimated numerically.

\end{appendix}
\bibliographystyle{unsrt}
\bibliography{Submission2}

\begin{thebibliography}{10}

\bibitem{hatzikirou2012go}
H.~Hatzikirou, D.~Basanta, M.~Simon, K.~Schaller, and A.~Deutsch.
\newblock ‘{G}o or grow’: the key to the emergence of invasion in tumour
  progression?
\newblock {\em Mathematical Medicine and Biology: A Journal of the IMA},
  29(1):49--65, 2012.

\bibitem{stepien2018traveling}
T.~L. Stepien, E.~M. Rutter, and Y.~Kuang.
\newblock Traveling waves of a go-or-grow model of glioma growth.
\newblock {\em SIAM Journal on Applied Mathematics}, 78(3):1778--1801, 2018.

\bibitem{crossley2023travelling}
R.~M. Crossley, P.~K. Maini, T.~Lorenzi, and R.~E. Baker.
\newblock Traveling waves in a coarse-grained model of volume-filling cell
  invasion: Simulations and comparisons.
\newblock {\em Studies in Applied Mathematics}, 151(4):1471--1497, 2023.

\bibitem{summerbell2020epigenetically}
E.~R. Summerbell, J.~K. Mouw, J.~S.~K. Bell, C.~M. Knippler, B.~Pedro, J.~L.
  Arnst, T.~O. Khatib, R.~Commander, B.~G. Barwick, J.~Konen, et~al.
\newblock Epigenetically heterogeneous tumor cells direct collective invasion
  through filopodia-driven fibronectin micropatterning.
\newblock {\em Science Advances}, 6(30):eaaz6197, 2020.

\bibitem{giese1996dichotomy}
A.~Giese, M.~A. Loo, N.~Tran, D.~Haskett, S.~W. Coons, and M.~E. Berens.
\newblock Dichotomy of astrocytoma migration and proliferation.
\newblock {\em International Journal of Cancer}, 67(2):275--282, 1996.

\bibitem{giese1996migration}
A.~Giese, L.~Kluwe, B.~Laube, H.~Meissner, M.~E. Berens, and M.~Westphal.
\newblock Migration of human glioma cells on myelin.
\newblock {\em Neurosurgery}, 38(4):755--764, 1996.

\bibitem{macfarlane2022impact}
F.~R. Macfarlane, T.~Lorenzi, and K.~J. Painter.
\newblock The impact of phenotypic heterogeneity on chemotactic
  self-organisation.
\newblock {\em Bulletin of Mathematical Biology}, 84(12):143, 2022.

\bibitem{lorenzi2022invasion}
T.~Lorenzi, B.~Perthame, and X.~Ruan.
\newblock Invasion fronts and adaptive dynamics in a model for the growth of
  cell populations with heterogeneous mobility.
\newblock {\em European Journal of Applied Mathematics}, 33(4):766--783, 2022.

\bibitem{fisher_wave_1937}
R.~A. Fisher.
\newblock {The} wave of advance of advantageous genes.
\newblock {\em Annals of Eugenics}, 7(4):355--369, 1937.

\bibitem{kolmogorov1937study}
A.~N. Kolmogorov, I.~Petrovskii, and N.~S. Piskunov.
\newblock A study of the equation of diffusion with increase in the quantity of
  matter, and its application to a biological problem.
\newblock {\em Moskovskogo University Bulletin of Mathematics}, 1(6):1--25,
  1937.

\bibitem{konen2017image}
J.~Konen, E.~Summerbell, B.~Dwivedi, K.~Galior, Y.~Hou, L.~Rusnak, A.~Chen,
  J.~Saltz, W.~Zhou, L.~H. Boise, et~al.
\newblock Image-guided genomics of phenotypically heterogeneous populations
  reveals vascular signalling during symbiotic collective cancer invasion.
\newblock {\em Nature Communications}, 8(1):15078, 2017.

\bibitem{campbell2021cooperation}
N.~R. Campbell, A.~Rao, M.~V. Hunter, M.~K. Sznurkowska, L.~Briker, M.~Zhang,
  M.~Baron, S.~Heilmann, M.~Deforet, C.~Kenny, et~al.
\newblock Cooperation between melanoma cell states promotes metastasis through
  heterotypic cluster formation.
\newblock {\em Developmental Cell}, 56(20):2808--2825, 2021.

\bibitem{kolbe2020modeling}
N.~Kolbe, N.~Sfakianakis, C.~Stinner, C.~Surulescu, and J.~Lenz.
\newblock Modeling multiple taxis: tumor invasion with phenotypic
  heterogeneity, haptotaxis, and unilateral interspecies repellence.
\newblock {\em Discrete and Continuous Dynamical Systems - B}, 26(1):443--481,
  2021.

\bibitem{pham2012density}
K.~Pham, A.~Chauviere, H.~Hatzikirou, X.~Li, H.~M. Byrne, V.~Cristini, and
  J.~Lowengrub.
\newblock Density-dependent quiescence in glioma invasion: instability in a
  simple reaction--diffusion model for the migration/proliferation dichotomy.
\newblock {\em Journal of Biological Dynamics}, 6(sup1):54--71, 2012.

\bibitem{gerlee2012impact}
P.~Gerlee and S.~Nelander.
\newblock The impact of phenotypic switching on glioblastoma growth and
  invasion.
\newblock {\em PLoS Computational Biology}, 8(6):e1002556, 2012.

\bibitem{gerlee2016travelling}
P.~Gerlee and S.~Nelander.
\newblock Travelling wave analysis of a mathematical model of glioblastoma
  growth.
\newblock {\em Mathematical Biosciences}, 276:75--81, 2016.

\bibitem{curtin2020speed}
L.~Curtin, A.~Hawkins-Daarud, K.~G. Van Der~Zee, K.~R. Swanson, and M.~R. Owen.
\newblock Speed switch in glioblastoma growth rate due to enhanced
  hypoxia-induced migration.
\newblock {\em Bulletin of Mathematical Biology}, 82(3):43, 2020.

\bibitem{tursynkozha2023traveling}
A.~Tursynkozha, A.~Kashkynbayev, B.~Shupeyeva, E.~M. Rutter, and Y.~Kuang.
\newblock Traveling wave speed and profile of a “go or grow” glioblastoma
  multiforme model.
\newblock {\em Communications in Nonlinear Science and Numerical Simulation},
  118:107008, 2023.

\bibitem{conte2021mathematical}
M.~Conte and C.~Surulescu.
\newblock Mathematical modeling of glioma invasion: acid-and vasculature
  mediated go-or-grow dichotomy and the influence of tissue anisotropy.
\newblock {\em Applied Mathematics and Computation}, 407:126305, 2021.

\bibitem{werb1997ecm}
Z.~Werb.
\newblock {ECM} and cell surface proteolysis: regulating cellular ecology.
\newblock {\em Cell}, 91(4):439--442, 1997.

\bibitem{cox2006new}
B.~D. Cox, M.~Natarajan, M.~R. Stettner, and C.~L. Gladson.
\newblock New concepts regarding focal adhesion kinase promotion of cell
  migration and proliferation.
\newblock {\em Journal of Cellular Biochemistry}, 99(1):35--52, 2006.

\bibitem{bloom2014influence}
A.~B. Bloom and M.~H. Zaman.
\newblock Influence of the microenvironment on cell fate determination and
  migration.
\newblock {\em Physiological Genomics}, 46(9):309--314, 2014.

\bibitem{zoeller2019genetic}
E.~L. Zoeller, B.~Pedro, J.~Konen, B.~Dwivedi, M.~Rupji, N.~Sundararaman,
  L.~Wang, J.~R. Horton, C.~Zhong, B.~G. Barwick, et~al.
\newblock Genetic heterogeneity within collective invasion packs drives leader
  and follower cell phenotypes.
\newblock {\em Journal of Cell Science}, 132(19):jcs231514, 2019.

\bibitem{mclennan2015neural}
R.~McLennan, L.~J. Schumacher, J.~A. Morrison, J.~M. Teddy, D.~A. Ridenour,
  A.~C. Box, C.~L. Semerad, H.~Li, W.~McDowell, D.~Kay, et~al.
\newblock Neural crest migration is driven by a few trailblazer cells with a
  unique molecular signature narrowly confined to the invasive front.
\newblock {\em Development}, 142(11):2014--2025, 2015.

\bibitem{vittadello2020examining}
S.~T. Vittadello, S.~W. McCue, G.~Gunasingh, N.~K. Haass, and M.~J. Simpson.
\newblock Examining go-or-grow using fluorescent cell-cycle indicators and
  cell-cycle-inhibiting drugs.
\newblock {\em Biophysical Journal}, 118(6):1243--1247, 2020.

\bibitem{GARAY20133094}
T.~Garay, E.~Juhász, E.~Molnár, M.~Eisenbauer, A.~Czirók, B.~Dekan,
  V.~László, M.~A. Hoda, B.~Döme, J.~Tímár, W.~Klepetko, W.~Walter~Berger,
  and B.~Hegedűs.
\newblock Cell migration or cytokinesis and proliferation? – revisiting the
  “go or grow” hypothesis in cancer cells in vitro.
\newblock {\em Experimental Cell Research}, 319(20):3094--3103, 2013.

\bibitem{alberts2017molecular}
B.~Alberts.
\newblock {\em Molecular Biology of the Cell}.
\newblock Garland Science, 2017.

\bibitem{gerisch2008mathematical}
A.~Gerisch and M.~A.~J. Chaplain.
\newblock Mathematical modelling of cancer cell invasion of tissue: local and
  non-local models and the effect of adhesion.
\newblock {\em Journal of Theoretical Biology}, 250(4):684--704, 2008.

\bibitem{browning2019bayesian}
A.~P. Browning, P.~Haridas, and M.~J. Simpson.
\newblock A {B}ayesian sequential learning framework to parameterise continuum
  models of melanoma invasion into human skin.
\newblock {\em Bulletin of Mathematical Biology}, 81(3):676--698, 2019.

\bibitem{lee2017local}
B.~Lee, J.~Konen, S.~Wilkinson, A.~I. Marcus, and Y.~Jiang.
\newblock Local alignment vectors reveal cancer cell-induced {ECM} fiber
  remodeling dynamics.
\newblock {\em Scientific Reports}, 7(1):39498, 2017.

\bibitem{winkler2020concepts}
J.~Winkler, A.~Abisoye-Ogunniyan, K.~J. Metcalf, and Z.~Werb.
\newblock Concepts of extracellular matrix remodelling in tumour progression
  and metastasis.
\newblock {\em Nature Communications}, 11(1):5120, 2020.

\bibitem{stetler1993tumor}
W.~G. Stetler-Stevenson, S.~Aznavoorian, and L.~A. Liotta.
\newblock Tumor cell interactions with the extracellular matrix during invasion
  and metastasis.
\newblock {\em Annual Review of Cell Biology}, 9(1):541--573, 1993.

\bibitem{chaplain2005mathematical}
M.~A.~J. Chaplain and G.~Lolas.
\newblock Mathematical modelling of cancer cell invasion of tissue: The role of
  the urokinase plasminogen activation system.
\newblock {\em Mathematical Models and Methods in Applied Sciences},
  15(11):1685--1734, 2005.

\bibitem{kessenbrock2010matrix}
K.~Kessenbrock, V.~Plaks, and Z.~Werb.
\newblock Matrix metalloproteinases: regulators of the tumor microenvironment.
\newblock {\em Cell}, 141(1):52--67, 2010.

\bibitem{perumpanani1999extracellular}
A.~J. Perumpanani and H.~M. Byrne.
\newblock Extracellular matrix concentration exerts selection pressure on
  invasive cells.
\newblock {\em European Journal of Cancer}, 35(8):1274--1280, 1999.

\bibitem{painter2009modelling}
K.~J. Painter.
\newblock Modelling cell migration strategies in the extracellular matrix.
\newblock {\em Journal of Mathematical Biology}, 58:511--543, 2009.

\bibitem{el2021travelling}
M.~El-Hachem, S.~W. McCue, and M.~J. Simpson.
\newblock Travelling wave analysis of cellular invasion into surrounding
  tissues.
\newblock {\em Physica D: Nonlinear Phenomena}, 428:133026, 2021.

\bibitem{colson2021travelling}
C.~Colson, F.~S{\'a}nchez-Gardu{\~n}o, H.~M. Byrne, P.~K. Maini, and
  T.~Lorenzi.
\newblock Travelling-wave analysis of a model of tumour invasion with
  degenerate, cross-dependent diffusion.
\newblock {\em Proceedings of the Royal Society A}, 477(2256):20210593, 2021.

\bibitem{chauviere2010model}
A.~Chauviere, L.~Preziosi, and H.~M. Byrne.
\newblock A model of cell migration within the extracellular matrix based on a
  phenotypic switching mechanism.
\newblock {\em Mathematical Medicine and Biology: A Journal of the IMA},
  27(3):255--281, 2010.

\bibitem{saut2014multilayer}
O.~Saut, J.-B. Lagaert, T.~Colin, and H.~M. Fathallah-Shaykh.
\newblock A multilayer grow-or-go model for {GBM}: effects of invasive cells
  and anti-angiogenesis on growth.
\newblock {\em Bulletin of Mathematical Biology}, 76:2306--2333, 2014.

\bibitem{colson_travelling-wave_2021}
C.~Colson, F.~S{\'a}nchez-Gardu{\~n}o, H.~M. Byrne, P.~K. Maini, and
  T.~Lorenzi.
\newblock Travelling-wave analysis of a model of tumour invasion with
  degenerate, cross-dependent diffusion.
\newblock {\em Proceedings of the Royal Society A}, 477(2256):20210593, 2021.

\bibitem{giese2003cost}
A.~Giese, R.~Bjerkvig, M.~E. Berens, and M.~Westphal.
\newblock Cost of migration: invasion of malignant gliomas and implications for
  treatment.
\newblock {\em Journal of Clinical Oncology}, 21(8):1624--1636, 2003.

\bibitem{mclennan2012multiscale}
R.~McLennan, L.~Dyson, K.~W. Prather, J.~A. Morrison, R.~E. Baker, P.~K. Maini,
  and P.~M. Kulesa.
\newblock Multiscale mechanisms of cell migration during development: theory
  and experiment.
\newblock {\em Development}, 139(16):2935--2944, 2012.

\bibitem{martinson2023dynamic}
W.~D. Martinson, R.~McLennan, J.~M. Teddy, M.~C. McKinney, L.~A. Davidson,
  R.~E. Baker, H.~M. Byrne, P.~M. Kulesa, and P.~K. Maini.
\newblock Dynamic fibronectin assembly and remodeling by leader neural crest
  cells prevents jamming in collective cell migration.
\newblock {\em Elife}, 12:e83792, 2023.

\bibitem{carmeliet2011molecular}
P.~Carmeliet and R.~K. Jain.
\newblock Molecular mechanisms and clinical applications of angiogenesis.
\newblock {\em Nature}, 473(7347):298--307, 2011.

\bibitem{deregibus2007endothelial}
M.~C. Deregibus, V.~Cantaluppi, R.~Calogero, M.~Lo~Iacono, C.~Tetta,
  L.~Biancone, S.~Bruno, B.~Bussolati, and G.~Camussi.
\newblock Endothelial progenitor cell--derived microvesicles activate an
  angiogenic program in endothelial cells by a horizontal transfer of m{RNA}.
\newblock {\em Blood, The Journal of the American Society of Hematology},
  110(7):2440--2448, 2007.

\bibitem{vilchez2021decoding}
S.~A. Vilchez~Mercedes, F.~Bocci, H.~Levine, J.~N. Onuchic, M.~K. Jolly, and
  P.~K. Wong.
\newblock Decoding leader cells in collective cancer invasion.
\newblock {\em Nature Reviews Cancer}, 21(9):592--604, 2021.

\bibitem{tosh2002cells}
D.~Tosh and J.~M.~W. Slack.
\newblock How cells change their phenotype.
\newblock {\em Nature Reviews Molecular Cell Biology}, 3(3):187--194, 2002.

\bibitem{quesenberry2010cellular}
P.~J. Quesenberry and J.~M. Aliotta.
\newblock Cellular phenotype switching and microvesicles.
\newblock {\em Advanced Drug Delivery Reviews}, 62(12):1141--1148, 2010.

\bibitem{lorenzi2016interfaces}
T.~Lorenzi, A.~Lorz, and B.~Perthame.
\newblock On interfaces between cell populations with different mobilities.
\newblock {\em Kinetic and Related Models}, 10(1):299--311, 2017.

\bibitem{drasdo2012modeling}
D.~Drasdo and S.~Hoehme.
\newblock Modeling the impact of granular embedding media, and pulling versus
  pushing cells on growing cell clones.
\newblock {\em New Journal of Physics}, 14(5):055025, 2012.

\bibitem{anderson2006tumor}
A.~R.~A. Anderson, A.~M. Weaver, P.~T. Cummings, and V.~Quaranta.
\newblock Tumor morphology and phenotypic evolution driven by selective
  pressure from the microenvironment.
\newblock {\em Cell}, 127(5):905--915, 2006.

\bibitem{dickinson1993stochastic}
R.~B. Dickinson and R.~T. Tranquillo.
\newblock A stochastic model for adhesion-mediated cell random motility and
  haptotaxis.
\newblock {\em Journal of Mathematical Biology}, 31(6):563--600, 1993.

\bibitem{nusgens1984collagen}
B.~Nusgens, C.~Merrill, C.~Lapiere, and E.~Bell.
\newblock Collagen biosynthesis by cells in a tissue equivalent matrix in
  vitro.
\newblock {\em Collagen and related research}, 4(5):351--363, 1984.

\bibitem{weinberg1985regulation}
C.~B. Weinberg and E.~Bell.
\newblock Regulation of proliferation of bovine aortic endothelial cells,
  smooth muscle cells, and adventitial fibroblasts in collagen lattices.
\newblock {\em Journal of Cellular Physiology}, 122(3):410--414, 1985.

\bibitem{yoshizato1985growth}
K.~Yoshizato, T.~Taira, and N.~Yamamoto.
\newblock Growth inhibition of human fibroblasts by reconstituted collagen
  fibrils.
\newblock {\em Biomedical research}, 6(2):61--71, 1985.

\bibitem{penington2011building}
C.~J. Penington, B.~D. Hughes, and K.~A. Landman.
\newblock Building macroscale models from microscale probabilistic models: a
  general probabilistic approach for nonlinear diffusion and multispecies
  phenomena.
\newblock {\em Physical Review E}, 84(4):041120, 2011.

\bibitem{simpson2009multi}
M.~J. Simpson, K.~A. Landman, and B.~D. Hughes.
\newblock Multi-species simple exclusion processes.
\newblock {\em Physica A: Statistical Mechanics and its Applications},
  388(4):399--406, 2009.

\bibitem{strobl2020mix}
M.~A.~R. Strobl, A.~L. Krause, M.~Damaghi, R.~Gillies, A.~R.~A. Anderson, and
  P.~K. Maini.
\newblock Mix and match: phenotypic coexistence as a key facilitator of cancer
  invasion.
\newblock {\em Bulletin of Mathematical Biology}, 82:1--26, 2020.

\bibitem{morton2005numerical}
K.~W. Morton and D.~F. Mayers.
\newblock {\em Numerical solution of partial differential equations: an
  introduction}.
\newblock Cambridge University Press, 2005.

\bibitem{nicolson1984generation}
G.~L. Nicolson.
\newblock Generation of phenotypic diversity and progression in metastatic
  tumor cells.
\newblock {\em Cancer and Metastasis Reviews}, 3:25--42, 1984.

\end{thebibliography}

\section*{Supplementary Information} 

\setcounter{figure}{0}  
\renewcommand{\thefigure}{S\arabic{figure}}

\setcounter{table}{0}  
\renewcommand{\thetable}{S\arabic{table}}

\setcounter{equation}{0}  
\renewcommand{\theequation}{S\arabic{equation}}

\setcounter{section}{0}
\renewcommand{\thesection}{S\arabic{section}}

\section{Distinct switching rates in either direction}\label{supp:s12}
In this section, we investigate the system \eqref{intro_u1_ND}-\eqref{intro_m_ND} subject to {the} phenotypic switching functions listed in Table~\ref{tab:PStab2} with different switching rates in either direction, such that cells in phenotypic state 1 switch to phenotypic state 2 at rate $s_{12}\in\mathbb{R}_{+}$ and, equivalently, cells in phenotypic state 2 switch to phenotypic state 1 at a rate given by $s_{21}\in\mathbb{R}_{+}.$
Since there is limited evidence that distinctly different switching rates are biologically relevant \cite{nicolson1984generation}, we only briefly discuss some of the main results for $s_{12}\neq s_{21}$ in this section to demonstrate these parameters' impact on the model {solutions}.

\begin{table*}[htbp]
\begin{center}
\begin{tabular}{|c|c|c|}
    \hline
    Name & $\gamma_{12}(u_1, u_2, m)$ & $\gamma_{21}(u_1, u_2, m)$\\
    \hline\hline
    Constant switching & $s_{12}$ & $s_{21}$ \\
    ECM-dependent switching & $s_{12} (1-m)$ & $s_{21} m$ \\
    Space-dependent switching & $s_{12}(1-u_1-u_2-m)$ & $s_{21}(u_1+u_2+m)$ \\
    Cell-dependent switching & 
    $s_{12}(1-u_1-u_2)$ & $s_{21}(u_1+u_2)$ \\
    \hline
\end{tabular}
\end{center}
\caption{Table listing the phenotypic switching functions.}
\label{tab:PStab2}
\end{table*}

\begin{figure}[htbp]
    \centering
   \includegraphics[scale=0.7]{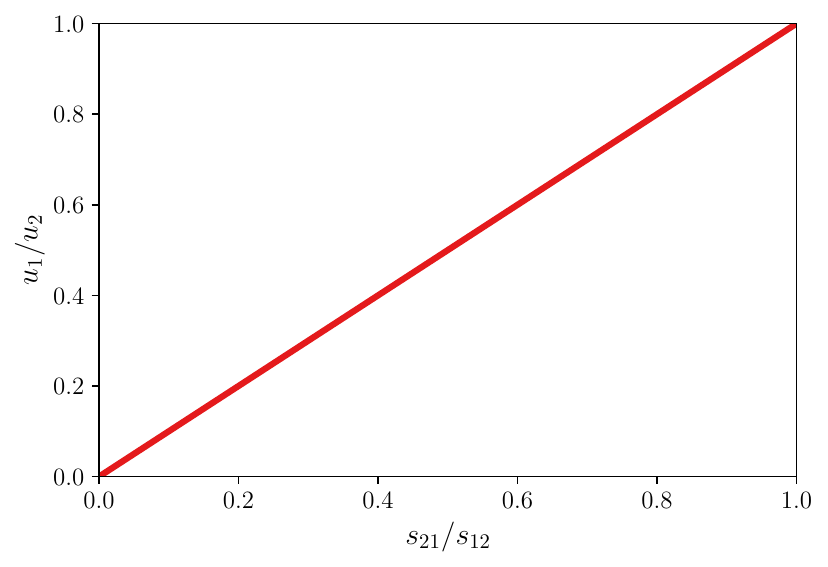}
    \caption{Plot demonstrating the relationship between the ratio of the switching rates, $s_{12}$ and $s_{21}$, and the ratio between the volume fraction of the two cell sub-populations behind the wave front, $u_1/u_2$ when simulating the system \eqref{intro_u1_ND}-\eqref{intro_m_ND} subject to the initial conditions for the cells as in Eq.~\eqref{IC_up}, and for the ECM as in Eq.~\eqref{IC_m}, subject to constant phenotypic switching (see Table~\ref{tab:PStab}). This plot was produced by running simulations under a variety of switching rates and plotting the ratio between the resulting cell sub-population densities behind the wave front. The initial ECM volume fraction ahead of the cells is $m_0=0.5$, the ECM degradation rate is $\lambda=1$, and the width of the region initially invaded by migrating cells is $\alpha=1$ across all simulations. For more information regarding the numerical methods used see~\ref{app_methods}.}
    \label{fig:CS_S_ratio}
\end{figure}

Fig.~\ref{fig:CS_S_ratio} shows that, in the case of constant switching, the ratio between the phenotypic switching rates, $s_{12}$ and $s_{21}$, directly determines the ratio between the volume fraction of cells in phenotypic states 1 and 2 in the bulk of the migrating cell population, as described in \ref{SHSS_CS}.  
For $s_{12},\, s_{21}\in[0,1]$ the relationship between the travelling wave speed and these parameters is symmetrical around $s_{12}=s_{21}$, which is the maximum speed {observed numerically and predicted analytically in the fast phenotypic switching regime (see Supplementary Material~\ref{supp:max})}. 

\begin{figure}
    \centering
    \includegraphics[width=\linewidth]{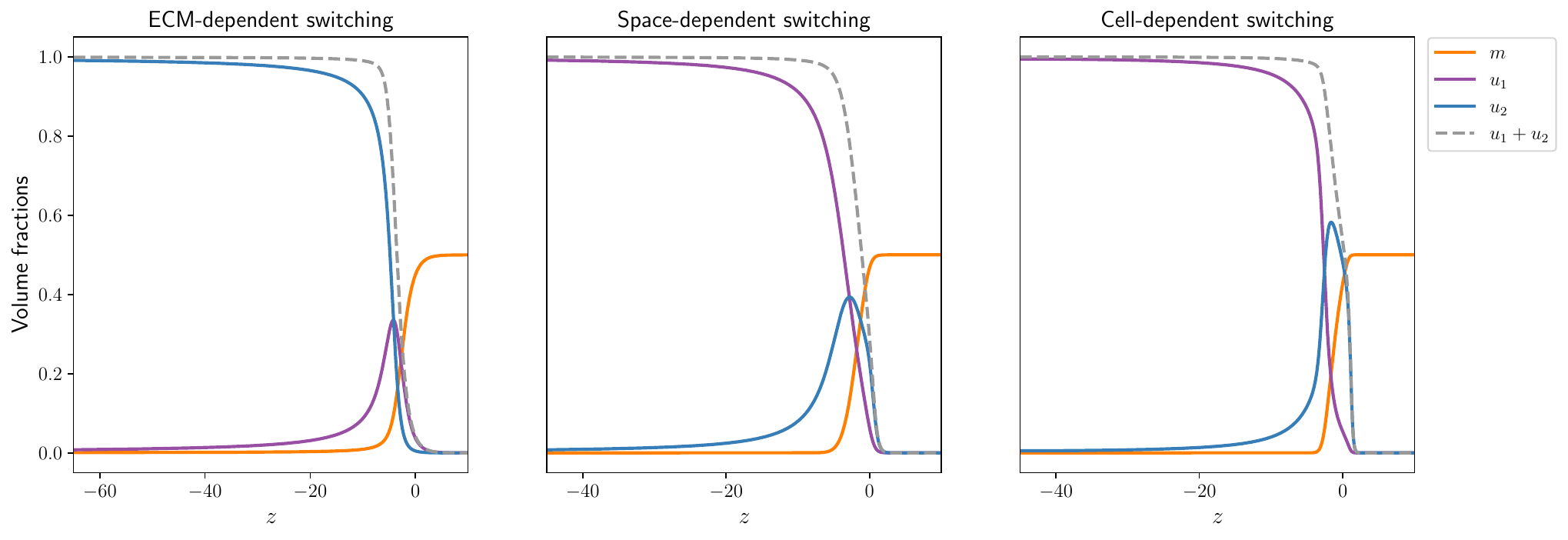}
    \caption{Travelling wave profiles of the solutions of Eqs.~\eqref{intro_u1_ND}-\eqref{intro_m_ND} subject to the initial conditions for the cells in Eq.~\eqref{IC_up}, and {for} the ECM in Eq.~\eqref{IC_m}{,} plotted {as a function of} the travelling wave variable $z=x-ct$, where $c$ is the numerically observed travelling wave speed{. These solutions} demonstrate that changing the switching rate in one direction leads to one {sub-population dominating} the migrating front. For ECM-dependent switching, the switching rate from phenotypic state 1 to 2 is $s_{12}=1$ and the switching rate from phenotypic state 2 to 1 is $s_{21}=10.$  For space- and cell-dependent switching, the switching rate from phenotypic state 1 to 2 is $s_{12}=10$ and the switching rate from phenotypic state 2 to 1 is $s_{21}=1.$ The initial ECM volume fraction ahead of the cells is $m_0=0.5$, the ECM degradation rate is $\lambda=1$, the weighting of the specialists towards degrading ECM is $\theta_{S,D}=0.5$ and the width of the region initially invaded by migrating cells is $\alpha=1$ in all cases.  For more information regarding the numerical methods used see \ref{app_methods}. }
    \label{fig:TWsep}
\end{figure}

Furthermore, changes in the individual switching rates also impact the travelling wave profile and speed of migration. 
For example, when considering ECM-dependent switching, although increasing the switching rate from phenotypic state 2 to phenotypic state 1 decreases the travelling wave speed, it also changes the distribution of cells in the migrating front such that the front of the travelling wave is dominated by degrading cells in phenotypic state 1, ahead of a mixed region of cells in both states, and the bulk of proliferating cells remains in the rear (see Fig.~\ref{fig:TWsep}). 

A similar result can be observed in Fig.~\ref{fig:TWsep} for space-dependent switching and cell-dependent switching. 
In these cases, increasing the switching rate from phenotypic state 1 to phenotypic state 2 causes a leading population of cells in phenotypic state 2 at the front of the travelling wave. 
In all cases, the greater the difference between the switching rates, the larger and more concentrated this proportion of leader cells are.

\section{{Formal t}ravelling wave analysis in the fast phenotypic switching regime}\label{supp:anasy}
In a regime where phenotypic switching is faster than cell motility and proliferation, we can consider the following rescaled model
\begin{align}
\frac{\partial {u_{1_\epsilon}}}{\partial {t}} &=(1-\theta_{S,D})\frac{\partial}{\partial {x}}\Bigg[ \big(1-u_{1_\epsilon}-u_{2_\epsilon}-m_\epsilon\big)\frac{\partial {u}_1}{\partial {x}}+{u_{1_\epsilon}} \frac{\partial }{\partial {x}}\big(u_{1_\epsilon}+u_{2_\epsilon}+m_\epsilon\big)\Bigg] \nonumber \\ &\qquad\qquad+\dfrac{1}{\epsilon}{{u}_{2_\epsilon}}{\gamma_{21}(u_{1_\epsilon}, u_{2_\epsilon}, m_\epsilon)}-\dfrac{1}{\epsilon}{u}_{1_\epsilon}{\gamma_{12}(u_{1_\epsilon}, u_{2_\epsilon}, m_\epsilon)},\nonumber \\
\frac{\partial {u_{2_\epsilon}}}{\partial {t}}&= \theta_{S,P}{{u}_{2_\epsilon}}\big(1-u_{1_\epsilon}-u_{2_\epsilon}-m_\epsilon\big)\nonumber \\ &\qquad\qquad-\dfrac{1}{\epsilon}{{u}_{2_\epsilon}}{\gamma_{21}(u_{1_\epsilon}, u_{2_\epsilon}, m_\epsilon)}+\dfrac{1}{\epsilon}{u}_{1_\epsilon}{\gamma_{12}(u_{1_\epsilon}, u_{2_\epsilon}, m_\epsilon)},  \label{intro_u2_NDe}\nonumber \\
\frac{\partial {m_\epsilon}}{\partial {t}}  &=-\theta_{S,D}\lambda {m}{u_{1_\epsilon}}, \nonumber
\end{align}
where $\epsilon\in\mathbb{R}_{+}$, $x\in\mathbb{R}$ and $t\in\mathbb{R}_{+}.$ 
{Here,} ${\lambda}\in\mathbb{R}_{+}$ is the rescaled rate of ECM degradation by cells in phenotypic state 1, whilst $\theta_{S,D}\in[0,1]$ describes the weighting of cells in phenotypic state 1 towards degrading ECM, $\theta_{S,P}=1$ describes the weighting of cells in phenotypic state 2 towards proliferation, and 
\begin{equation*}
    \gamma_{12}:\mathbb{R}_{+}^3\rightarrow\mathbb{R}_{+} \qquad \text{and} \qquad \gamma_{21}:\mathbb{R}_{+}^3\rightarrow\mathbb{R}_{+},
\end{equation*}
are the non-dimensional phenotypic switching functions.

Simulations reveal {that the model admits} {constant profile, constant speed} travelling wave solutions, so we introduce the travelling wave ansatz 
\begin{align}
    U_{p_\epsilon}(z) &= U_{p_\epsilon}(x-ct) = u_{p_\epsilon}(x,t), \nonumber\\
    M_\epsilon(z) &= M_\epsilon(x-ct)\,= m_\epsilon(x,t),\nonumber
\end{align}
for $p=\{1,\,2\},$ where $c\in\mathbb{R}_{+}, $ that satisfy the following system of ODEs:
\begin{align}
    -c \dfrac{\mathrm{d}U_{1_\epsilon}}{\mathrm{d}z}  &=(1-\theta_{S,D})\dfrac{\mathrm{d}}{\mathrm{d}z}\bigg[(1-U_{1_\epsilon}- U_{2_\epsilon}-M_\epsilon) \dfrac{\mathrm{d}U_{1_\epsilon}}{\mathrm{d}z}+U_{1_\epsilon}\dfrac{\mathrm{d}}{\mathrm{d}z}(U_{1_\epsilon}+U_{2_\epsilon}+M_\epsilon)\bigg]\nonumber \\ &\qquad\qquad -\dfrac{1}{\epsilon}{U}_{1_\epsilon}{\gamma_{12}(U_{1_\epsilon}, U_{2_\epsilon}, M_\epsilon)}+\dfrac{1}{\epsilon}{U}_{2_\epsilon}{\gamma_{21}(U_{1_\epsilon}, U_{2_\epsilon}, M_\epsilon)}, \label{epsU1} \\
    -c \dfrac{\mathrm{d}U_{2_\epsilon}}{\mathrm{d}z}  &= U_{2_\epsilon} (1-U_{1_\epsilon} - U_{2_\epsilon} - M_\epsilon) \nonumber \\ &\qquad\qquad +\dfrac{1}{\epsilon}{U}_{1_\epsilon}{\gamma_{12}(U_{1_\epsilon}, U_{2_\epsilon}, M_\epsilon)}-\dfrac{1}{\epsilon}{U}_{2_\epsilon}{\gamma_{21}(U_{1_\epsilon}, U_{2_\epsilon}, M_\epsilon)}, \label{epsU2}\\
    -c \dfrac{\mathrm{d}M_\epsilon}{\mathrm{d}z}  &= -\lambda \theta_{S, D} U_{1_\epsilon} M_\epsilon. \label{epsM}
\end{align}
Combining Eqs.~\eqref{epsU1}~and~\eqref{epsU2} we find that the total cell volume fraction $$U_\epsilon (z) = U_{1_\epsilon} (z) + U_{2_\epsilon} (z),$$ {satisfies} the ODE
\begin{equation}
     -c \dfrac{\mathrm{d}U_\epsilon}{\mathrm{d}z} = (1-\theta_{S,D})\dfrac{\mathrm{d}}{\mathrm{d}z}\bigg[(1-U_\epsilon-M_\epsilon) \dfrac{\mathrm{d}U_{1_\epsilon}}{\mathrm{d}z}+U_{1_\epsilon}\dfrac{\mathrm{d}}{\mathrm{d}z}(U_\epsilon+M_\epsilon)\bigg] + U_{2_\epsilon} (1-U_\epsilon - M_\epsilon), \label{totalUeps}
\end{equation}
for $z\in\mathbb{R}.$

Now consider constant phenotypic switching as defined in Table~\ref{tab:PStab2} and look for an analytical expression for the travelling wave speed in asymptotic regions of $\lambda$, following the ideas in \cite{crossley2023travelling}.

By considering the asymptotic expansions around $U_\epsilon, \, U_{1_\epsilon}, \, U_{2_\epsilon}$ and $M_\epsilon$ such that the leading-order terms are given by $U, \,  U_1, \, U_2$ and $M$, respectively, then as $\epsilon\to0^{+}$ we {formally} find, from Eqs.~\eqref{epsU1}~and~\eqref{epsU2}, that
\begin{equation} U_p(z)= \omega_p(U_1, U_2, M) U, \end{equation}
where 
\begin{align}
    \omega_1 &= \dfrac{s_{21}}{s_{12}+s_{21}},\nonumber \\
    \omega_2 &= \dfrac{s_{12}}{s_{12}+s_{21}}.\nonumber 
\end{align}
By substitution, Eq.~\eqref{totalUeps} becomes 
\begin{equation}
     -c \dfrac{\mathrm{d}U}{\mathrm{d}z} =\omega_1(1-\theta_{S,D})\dfrac{\mathrm{d}}{\mathrm{d}z}\bigg[(1-U-M) \dfrac{\mathrm{d}U}{\mathrm{d}z}+U\dfrac{\mathrm{d}}{\mathrm{d}z}(U+M)\bigg] + \omega_2 U (1-U - M), \label{total_U_LO}\nonumber
\end{equation}
which can be expanded and written as 
\begin{equation}
    \omega_1(1-\theta_{S,D})(1-M)\dfrac{\mathrm{d}^2 U}{\mathrm{d}z^2}+c\dfrac{\mathrm{d}U}{\mathrm{d}z}+\omega_2U(1-U-M)=-\omega_1 (1-\theta_{S,D})U\dfrac{\mathrm{d}^2M}{\mathrm{d}z^2}. \label{UU}
\end{equation}
Furthermore, Eq.~\eqref{epsM} can be written {as}
\begin{equation}
    c \dfrac{\mathrm{d}M}{\mathrm{d}z} = \lambda \theta_{S, D} \omega_1 U M, \label{M}
\end{equation}
which yields
\begin{align}
    \dfrac{\mathrm{d}^2M}{\mathrm{d}z^2} &= \dfrac{\lambda \theta_{S,D}\omega_1}{c}\bigg(U\dfrac{\mathrm{d}M}{\mathrm{d}z}+M\dfrac{\mathrm{d}U}{\mathrm{d}z}\bigg) \nonumber \\
    &=\dfrac{\lambda \theta_{S,D}\omega_1}{c}\bigg(\dfrac{\lambda\theta_{S, D}\omega_1}{c}MU^2+M\dfrac{\mathrm{d}U}{\mathrm{d}z}\bigg). \label{M2}
\end{align}
Substituting Eq.~\eqref{M2} into Eq.~\eqref{UU} we find
\begin{align}
    &\omega_1(1-\theta_{S,D})(1-M)\dfrac{\mathrm{d}^2 U}{\mathrm{d}z^2}+c\dfrac{\mathrm{d}U}{\mathrm{d}z}+\omega_2U(1-U-M)\nonumber \\ &\qquad \qquad=-\dfrac{\lambda \theta_{S,D}\omega_1^2 (1-\theta_{S,D})}{c}M U\bigg[\dfrac{\lambda \theta_{S,D}\omega_1}{c}U^2+\dfrac{\mathrm{d}U}{\mathrm{d}z}\bigg]. \label{UU2}
\end{align}
Moreover, solving Eq.~\eqref{M} subject to the boundary condition $M(z)\to m_0$ as $z\to\infty$, where $m_0\in[0,1],$ gives
\begin{equation}
    M(z) = m_0 \text{exp}\bigg\{ - \dfrac{\lambda \theta_{S,D}\omega_1}{c} \int_{z}^{\infty}U(s) \mathrm{d}s\bigg\}. 
\end{equation}
Under the boundary conditions $U_p(z)\to0$ as $z\to\infty $ for $p=1,\,2$ we have $U(z)\to0$ as $z\to\infty$. 
At the migrating front of the travelling wave (i.e., for $z\in(\ell, \infty)$ with $1\ll \ell < \infty$), we can use the ansatz 
\begin{equation}
    U(z)\approx \text{exp}\big\{-\beta z\big\},\nonumber
\end{equation}
where $\beta\in(0, \infty)$ to give 
\begin{equation}
    M(z) = m_0 \text{exp}\bigg\{ - \dfrac{\lambda \theta_{S,D}\omega_1}{\beta c} U(z) \bigg\}, \label{MU}
\end{equation}
for $z\in(\ell, \infty).$

\subsection{Formal asymptotic analysis for $\lambda \to 0 ^{+}$}
\begin{figure}[htbp]
    \centering
    \includegraphics[scale=0.6]{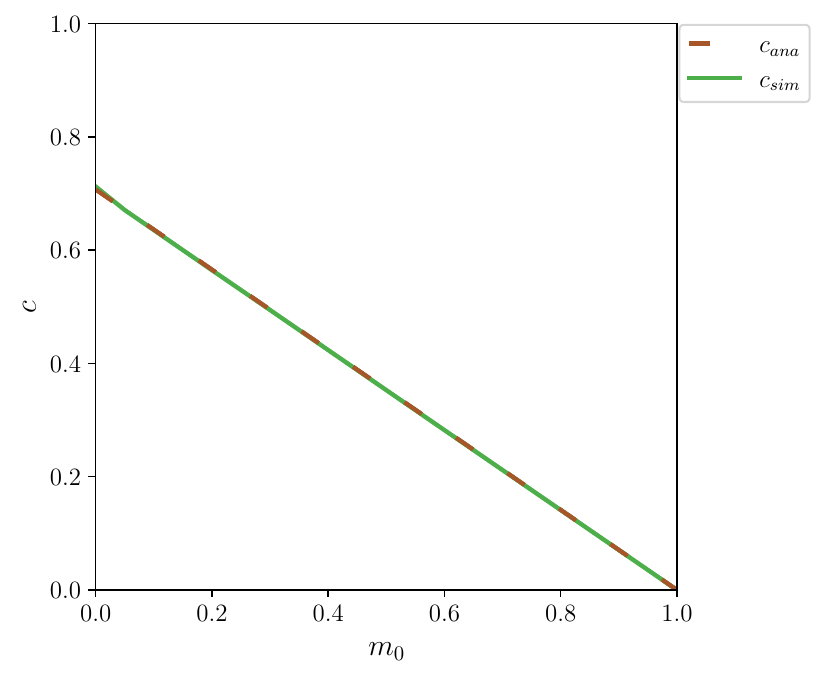}
    \caption{Plot showing the {analytically predicted minimum travelling wave speed, $c_{\text{ana}}$, and the numerically estimated} travelling wave speed{, $c_{\text{sim}}$,} of solutions of the system \eqref{intro_u1_ND}-\eqref{intro_m_ND} subject to the initial conditions for the cells as in Eq.~\eqref{IC_up}, and {for} the ECM as in Eq.~\eqref{IC_m} for very low ECM degradation rates, $\lambda\to0^{+}$, and various initial ECM volume fractions ahead of the cells, $m_0$, in fast phenotypic switching regimes, defined by simulations with $s_{12}=s_{21}=s=10^{4}$. The numerically estimated travelling wave speeds plotted in green are for simulations with $\lambda\leq10^{-2}$, while the analytical wave speeds plotted in red are given by Eq.~\eqref{clow}. The width of the {region initially invaded by} migrating cells is $\alpha=1$ and the weighting of specialists towards degradation is $\theta_{S,D}=0.5$ across all simulations. For more information regarding the numerical methods used see \ref{app_methods}.}
    \label{fig:fastPS_c}
\end{figure}
Using Eq.~\eqref{MU}, it is clear that 
\begin{equation}
    M(z) \approx m_0 \text{exp}\bigg\{ - \dfrac{\lambda \theta_{S,D}\omega_1}{\beta c} U(z) \bigg\} \to m_0 \qquad \text{as} \qquad \lambda\to0^{+}, \label{M0}
\end{equation}
for $z\in(\ell, \infty).$
In the asymptotic regime $\lambda \to 0^{+},$ since $0< U(z)< 1$ and $\mathrm{d}U(z)/\mathrm{d}z\approx -\beta U(z)$ for $z\in(\ell, \infty)$, substituting Eq.~\eqref{M0} into Eq.~\eqref{UU2} {results in} the asymptotic relation
\begin{equation}
    U(z) m_0 \text{exp}\bigg\{ - \dfrac{\lambda \theta_{S,D}\omega_1}{\beta c} U(z) \bigg\}\bigg[\dfrac{\lambda \theta_{S,D}\omega_1}{c}U^2(z)+\dfrac{\mathrm{d}U(z)}{\mathrm{d}z}\bigg]\to0 \qquad \text{as} \qquad \lambda\to0^{+},\nonumber
\end{equation}
for $z\in(\ell, \infty).$ Formally, we find
\begin{equation}
    \omega_1(1-\theta_{S,D})(1-M)\dfrac{\mathrm{d}^2 U}{\mathrm{d}z^2}+c\dfrac{\mathrm{d}U}{\mathrm{d}z}+\omega_2U(1-U-M)\approx 0, \label{F1}
\end{equation}
for $z\in(\ell, \infty).$ 
We notice that Eq.~\eqref{F1} is equivalent to the Fisher-KPP model \cite{fisher_wave_1937, kolmogorov1937study} in travelling-wave co-ordinates:
\begin{equation}
    \tilde{D}\dfrac{\mathrm{d}^2\tilde{U}(z)}{\mathrm{d}z^2} +\tilde{c}\dfrac{\mathrm{d}\tilde{U}}{\mathrm{d}z}+\tilde{r}\tilde{U}\bigg(1-\dfrac{\tilde{U}}{\tilde{K}}\bigg)=0, \label{FKPP}
\end{equation}
where we have $\tilde{D}=\omega_1(1-\theta_{S,D})(1-m_0), \, \tilde{r}=\omega_2(1-m_0)$ and $\tilde{K}=(1-m_0)$. 
This correctly predicts (see Fig.~\ref{fig:fastPS_c}), as $\lambda\to 0^{+}$, a minimum travelling wave speed given by 
\begin{equation}
    c_{\text{min}}=2(1-m_0)\sqrt{\omega_1\omega_2(1-\theta_{S,D})}, \label{clow}
\end{equation}
which can be observed in Fig.~\ref{fig:fastPS_c} to {agree with the numerically estimated wave speed for} the system \eqref{intro_u1_ND}-\eqref{intro_m_ND} {when} $\lambda\to0^{+}.$

\subsection{Formal asymptotic analysis for $\lambda \to \infty$}
\begin{figure}[htbp]
    \centering
    \includegraphics[scale=0.6]{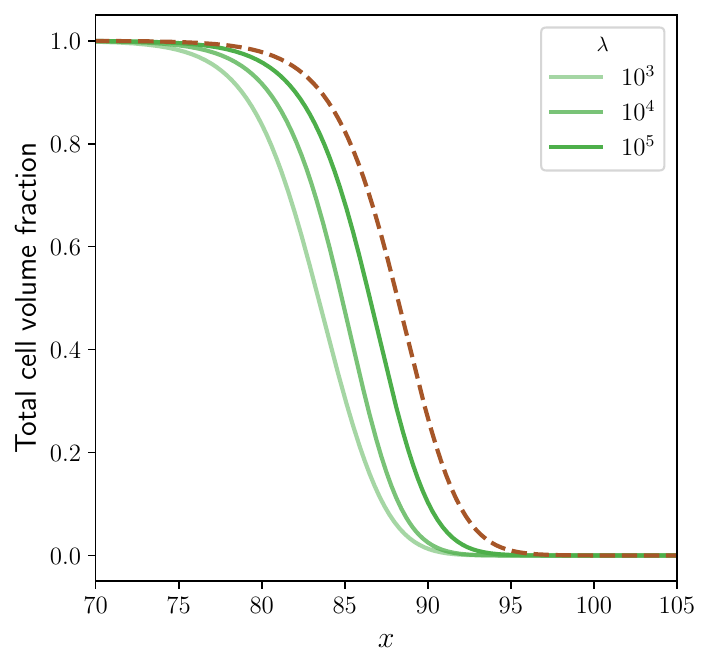}
    \caption{{Plot of the total cell volume fraction obtained through numerical simulations of Eqs.~\eqref{intro_u1_ND}-\eqref{intro_m_ND} with constant phenotypic switching (see Table~\ref{tab:PStab}) subject to the initial conditions for the cells as in Eq.~\eqref{IC_up} and {for} the ECM as in Eq.~\eqref{IC_m} for large values of $\lambda$ (solid lines), and numerical simulations of the Fisher-KPP model given by Eq.~\eqref{F2} (dashed red line). In all simulations, solutions are shown at $t=100$ and the initial ECM volume fraction ahead of the cells is $m_0=0.1$. The width of {the region initially invaded by} migrating cells is $\alpha=1$, the weighting of specialists towards degradation is $\theta_{S,D}=0.1$ and the switching rate for all functions is $s=\lambda$.  Qualitatively, the same behaviour is observed for all $m_0\in[0, 1)$. For more information regarding the numerical methods used see \ref{app_methods}.}}
    \label{fig:fastPS_c2}
\end{figure}

Revisiting the semi-explicit solution for $M$ given by Eq.~\eqref{MU}, we find
\begin{equation}
    M(z) \approx m_0 \text{exp}\bigg\{ - \dfrac{\lambda \theta_{S,D}\omega_1}{\beta c} U(z) \bigg\} \to 0 \qquad \text{as} \qquad \lambda\to\infty, \label{Minf}
\end{equation}
for $z\in(\ell, \infty).$
In the asymptotic regime $\lambda \to \infty${,} since $0< U(z)< 1$ and $\mathrm{d}U(z)/\mathrm{d}z\approx -\beta U(z)$ for $z\in(\ell, \infty)${,} substituting Eq.~\eqref{Minf} into Eq.~\eqref{UU} {results in} the asymptotic relation
\begin{equation}
    U(z) m_0 \text{exp}\bigg\{ - \dfrac{\lambda \theta_{S,D}\omega_1}{\beta c} U(z) \bigg\}\bigg[\dfrac{\lambda \theta_{S,D}\omega_1}{c}U^2(z)+\dfrac{\mathrm{d}U(z)}{\mathrm{d}z}\bigg]\to0 \qquad \text{as} \qquad \lambda\to\infty,\nonumber
\end{equation}
for $z\in(\ell, \infty).$
By substitution, we then find
\begin{equation}
    \omega_1(1-\theta_{S,D})\dfrac{\mathrm{d}^2 U}{\mathrm{d}z^2}+c\dfrac{\mathrm{d}U}{\mathrm{d}z}+\omega_2U(1-U)\approx 0, \label{F2}
\end{equation}
for $z\in(\ell, \infty).$ 
In this case, Eq.~\eqref{F2} is equivalent to the Fisher-KPP model (see Eq.~\eqref{FKPP}) with parameters $\tilde{D}=\omega_1(1-\theta_{S,D}), \, \tilde{r}=\omega_2$ and $\tilde{K}=1$, so when $\lambda\to\infty$ we have
\begin{equation}
    c_{\text{min}}=2\sqrt{\omega_1\omega_2(1-\theta_{S,D})}. \label{chigh}
\end{equation}
Fig.~\ref{fig:fastPS_c2} shows the convergence of the solutions to the system \eqref{intro_u1_ND}-\eqref{intro_m_ND} to the solution of the Fisher-KPP model (see Eq.~\eqref{FKPP}) with parameters $\tilde{D}=\omega_1(1-\theta_{S,D}), \, \tilde{r}=\omega_2$ and $\tilde{K}=1$ as $\lambda\to\infty.$

\subsection{Maximising the travelling wave speed}\label{supp:max}
In both $\lambda\to0^{+}$ and $\lambda\to\infty$ regimes, the travelling wave speeds, determined by Eq.~\eqref{clow} and Eq.~\eqref{chigh}{,} respectively, are maximised when $\omega_1\omega_2$ is maximised.
By considering 
\begin{equation}
    \omega_1\omega_2 =  \dfrac{s_{12}s_{21}}{(s_{12}+s_{21})^2}, \label{oms}
\end{equation}
and differentiating twice with respect to $s_{12}$, it is clear that 
\begin{equation}
    \text{max}(\omega_1\omega_2)=0.25,
\end{equation}
which is obtained when $s_{12}=s_{21}.$
As such, we can conclude that the travelling wave speed is always maximised when phenotypic switching between states 1 and 2 occur{s} at the same rate in either direction.

\end{document}